\documentclass[final,5p,twocolumn,authoryear]{elsarticle}

\usepackage[utf8]{inputenc}
\usepackage[T1]{fontenc}
\usepackage{amsmath}
\usepackage{newtxtext,newtxmath}
\usepackage{graphicx,array,booktabs}
\usepackage{float}
\usepackage{caption}
\usepackage{placeins}
\usepackage{stfloats}


\newenvironment{fwfigure}[1][!tbp]{\begin{figure*}[#1]}{\end{figure*}}
\newenvironment{fwtable}[1][!tbp]{\begin{table*}[#1]}{\end{table*}}
\usepackage{microtype}
\usepackage[mathlines,switch]{lineno}
\usepackage{hyperref}
\hypersetup{hidelinks,
  pdftitle={Closed-loop policy optimisation for cylinder wake control via differentiable simulation},
  pdfauthor={Siwen Zhang, Xizeng Zhao, Zhouteng Ye, Yulin Xie, Zesheng Yao},
  pdfsubject={Active flow control},
  pdfkeywords={active flow control, differentiable simulation, policy optimisation, vortex shedding suppression, synthetic jets}}
\numberwithin{equation}{section}
\newcounter{algorithm}

\newcommand{\Real}{\mathbb{R}}
\newcommand{\E}{\mathbb{E}}

\newcommand{\Cd}{C_d}
\newcommand{\Cl}{C_l}
\newcommand{\Rey}{Re}
\newcommand{\bfu}{\boldsymbol{u}}

\newcommand{\bfs}{\boldsymbol{s}}
\newcommand{\bfo}{\boldsymbol{o}}
\newcommand{\bfx}{\boldsymbol{x}}
\newcommand{\loss}{\mathcal{L}}   
\newcommand{\Loss}{\mathrm{L}}    

\journal{Ocean Engineering}

\begin{document}

\begin{frontmatter}

\title{Closed-loop policy optimisation for cylinder wake control via differentiable simulation}

\author[1]{Siwen Zhang}
\ead{12334047@zju.edu.cn}

\author[1,2]{Xizeng Zhao\corref{cor1}}
\ead{xizengzhao@zju.edu.cn}

\author[3]{Zhouteng Ye\corref{cor2}}
\ead{zhoutengye@buaa.edu.cn}

\author[4]{Yulin Xie}
\ead{bsh038@csust.edu.cn}

\author[5]{Zesheng Yao}
\ead{2246378226@qq.com}

\affiliation[1]{organization={Ocean College, Zhejiang University},
  city={Zhoushan}, postcode={316021}, state={Zhejiang}, country={China}}
\affiliation[2]{organization={Ocean Research Center of Zhoushan, Zhejiang University},
  city={Zhoushan}, postcode={316021}, state={Zhejiang}, country={China}}
\affiliation[3]{organization={Hangzhou International Innovation Institute, Beihang University},
  city={Hangzhou}, postcode={311115}, state={Zhejiang}, country={China}}
\affiliation[4]{organization={School of Hydraulic and Ocean Engineering, Changsha University of Science \& Technology},
  city={Changsha}, postcode={410114}, state={Hunan}, country={China}}
\affiliation[5]{organization={The State Key Laboratory of Fluid Power \& Mechatronic Systems, Zhejiang~University},
  city={Hangzhou}, postcode={310058}, state={Zhejiang}, country={China}}

\cortext[cor1]{Corresponding author.}
\cortext[cor2]{Co-corresponding author.}

\begin{abstract}

Suppressing vortex shedding behind a circular cylinder is a classical problem in active flow control.
Model-free deep reinforcement learning (DRL) learns such policies by interacting with the flow without differentiating its dynamics, which typically requires a large interaction budget. 
This study optimises closed-loop policies by differentiable simulation: control-loss gradients propagate through the simulated flow to the policy parameters, which are updated by reverse-mode automatic differentiation. 
The policy maps local velocity measurements to a jet amplitude imposed through differentiable synthetic-jet boundary conditions. 
In a two-dimensional confined cylinder wake at $\Rey=100$, eight policy updates (about 78 shedding periods of forward simulation) give approximately 8\% drag reduction and 99.6\% suppression of lift fluctuations, the drag level being comparable to reported DRL results. 
Suppression is sustained by feedback: a mean-matched constant jet does not maintain it, whereas the same policy responds similarly at three activation phases and recovers after upstream disturbances. 
A velocity-deviation reward alone retains strong suppression, and observations of two instantaneous velocity components at one near-wake or surface location reach about 99.8\% suppression, whereas a single surface pressure leaves residual oscillations. 
Together with related DRL results, this suggests that reward and observation requirements here are set by policy optimisation and flow physics.


\end{abstract}

\begin{keyword}
active flow control \sep differentiable simulation \sep policy optimisation \sep vortex shedding suppression \sep synthetic jets
\end{keyword}

\end{frontmatter}


\section{Introduction}\label{sec:introduction}

Flow past a circular cylinder develops periodic vortex shedding
once the steady wake becomes unstable. The alternating vortices produce
lift fluctuations and increase mean drag
\citep{Williamson1996,Choi2008}. Suppressing these oscillations is a
classical problem in active flow control (AFC). The fixed-cylinder wake
combines a simple geometry with unsteady vortex dynamics and measurable
force responses, allowing the effects of actuation on wake structures and
hydrodynamic loads to be examined together \citep{BruntonNoack2015,Rabault2019}.

AFC modifies the flow through external actuation
\citep{Gad-el-Hak2000,Cattafesta2011}, with synthetic jets providing a means
of manipulating bluff-body wakes \citep{Choi2008,Glezer2011}.
Designing the jet input is challenging because the wake dynamics are
nonlinear and unsteady, and each control action influences the subsequent
flow evolution \citep{Sipp2016,Kim2003,BruntonNoack2015}.
Effective control must reduce the loads and maintain the suppressed wake
state \citep{Li2022Confined}. Closed-loop control uses flow measurements
to adjust the jet actuation as the flow evolves, suppressing oscillations
and maintaining the suppressed state.

Machine learning offers a way to construct this observation-to-action
mapping, including through deep reinforcement learning (DRL)
\citep{Brunton2020,Garnier2021,VinuesaBrunton2022,Vinuesa2024}.
In numerical AFC, model-free DRL learns from trajectories and returns
generated by forward flow simulations, without requiring derivatives of
the flow solver. Generating these trajectories accounts for a substantial
part of the training cost, motivating parallel training environments
\citep{Garnier2021,Rabault2020}.
For a cylinder wake at $\Rey=100$, \citet{Rabault2019} used proximal
policy optimisation (PPO, \citealt{Schulman2017}) to train an artificial
neural network (ANN) for synthetic-jet control from velocity observations.
Subsequent developments include the DRLinFluids platform
\citep{Wang2022DRLinFluids}, control across Reynolds numbers
\citep{Tang2020}, and applications to weakly turbulent wakes and turbulent
channel flows \citep{Ren2021,Sonoda2023}.
Control performance also depends on observation and reward design.
Studies of sparse sensing and probe selection have examined the role of
observations \citep{Castellanos2022,Paris2021}.
\citet{Li2022Confined} used flow sensitivity and stability information to
guide probe placement and reward design for confined-cylinder wakes.
\citet{Wang2024Dynamic} used pressure and action histories to improve
circular-cylinder control with sparse surface-pressure measurements.
\citet{Xia2024} incorporated measurement and action histories for
square-cylinder control under partial observations. The importance of
reward and observation design in these studies raises the question of
whether direct flow-response gradients could support simpler designs
while retaining effective suppression.

Differentiable simulation provides gradients of the control objective
with respect to actuation through the numerical flow evolution
\citep{Holl2020}. Automatic differentiation (AD) computes these derivatives
by applying the chain rule to the implemented numerical operations
\citep{Griewank2008}. Adjoint methods have long used objective gradients
for aerodynamic design and flow control \citep{Jameson1988,Bewley2001},
and reverse-mode AD shares the chain-rule structure of discrete adjoints
\citep{Sirkes1997,Zhu2021}. AD can differentiate the solver trajectory
directly or supply Jacobians for a separately solved adjoint equation
\citep{SirignanoMacArtFreund2020,LiuMacArt2024}.
Differentiable solvers such as $\Phi_\textrm{Flow}$ (PhiFlow)
\citep{Holl2020,Holl2024PhiFlow} support this computation, while related
solver-in-the-loop approaches have trained flow corrections and turbulence
models \citep{Um2020,Kochkov2021,List2022}.
When a neural network supplies the control input, the gradients can
propagate further to its parameters. The policy can then be updated
directly from the simulated flow response, providing an alternative to
policy updates estimated from sampled returns in model-free DRL
\citep{Holl2020,Schulman2017}.

Existing gradient-based flow studies optimise several types of variables,
including actuation fields, design parameters, action sequences and
neural-network controller parameters (table~\ref{tab:lit-comparison}).
\citet{WangChu2025} optimised wall actuation and permeability fields in
compressible turbulent channel flows using a receding-horizon procedure.
\citet{Alhashim2025} applied end-to-end AD to fluid-mechanical inverse
design, while differentiable model predictive control in FluidGym
optimises action sequences through online flow prediction
\citep{Becktepe2026FluidGym}. \citet{Holl2020} trained prediction and control networks
through a differentiable solver within a predictor--corrector
scheme, combining a trajectory-prediction
network with a control network that maps current observations and predicted
target states to body forces. \citet{LiuMacArt2024} used AD-assisted adjoint
calculations to optimise a neural controller mapping local pressure to
body forces in cylinder flow. \citet{LiuHicklingMacArt2025} used
adjoint-based gradients to train a neural controller mapping local surface
pressure to jet total pressure in airfoil flow. For synthetic-jet control of the
confined-cylinder wake, a neural policy can set the jet velocity through
differentiable boundary conditions. Gradients of the control objective
can then propagate through the simulated flow response to update the
network parameters. Applying this approach to the same flow-control
problem studied with DRL provides a reference for comparing cumulative
training interactions and suppression performance for policies driven by
local measurements. Building on the reward and observation designs
explored in DRL, we ask whether direct flow-response gradients support
simpler designs while retaining effective suppression in this wake.

\begin{fwtable}[tbp]
  \centering
  \caption{Classification of representative gradient-based studies by
    problem setting and optimisation approach.
    AD-adjoint denotes adjoint gradient calculations assisted by AD. D-MPC denotes differentiable model predictive control
    (online action-sequence optimisation).
    An optimised ANN policy maps observations to actions without online
    trajectory optimisation.}
  \label{tab:lit-comparison}
  {\small
  \begin{tabular}{@{} >{\raggedright\arraybackslash}p{0.20\textwidth} >{\raggedright\arraybackslash}p{0.42\textwidth} >{\raggedright\arraybackslash}p{0.30\textwidth} @{}}
    \hline
    Study & Problem setting & Optimisation approach \\
    \hline
    \citet{WangChu2025}           & Channel AFC (compressible turbulent); opposition control
                                  & End-to-end AD; receding-horizon control \\
    \citet{Alhashim2025}          & Inverse design (various)
                                  & End-to-end AD; parameter optimisation \\
    \citet{Becktepe2026FluidGym}  & \mbox{Multi-environment} AFC; jet\,/\,rotation\,/\,blowing
                                  & End-to-end AD (D-MPC)\,/\,DRL \\[4pt]
    \citet{Holl2020}              & Abstract PDE control; body force
                                  & End-to-end AD; prediction and control networks \\
    \citet{LiuMacArt2024}         & Bluff-body AFC, circular cylinder; body force
                                  & AD-adjoint + ANN policy \\
    \citet{LiuHicklingMacArt2025} & Airfoil AFC (compressible); jet
                                  & AD-adjoint + ANN policy \\
    \hline
  \end{tabular}}
\end{fwtable}

The present work applies differentiable simulation to neural-network
policy optimisation for a two-dimensional confined circular-cylinder wake
at $\Rey=100$, controlled by symmetrically positioned synthetic jets.
We couple the policy to a PhiFlow simulation through differentiable jet
velocity boundary conditions and propagate control-objective gradients
to the network parameters. We will assess training progress and compare
cumulative training interactions, drag reduction and lift-fluctuation
suppression with published DRL results.
We will examine sustained suppression, responses to different starting
phases and recovery after upstream disturbances with the conventional
force-coefficient reward. Separate comparisons will assess a simpler
velocity-based reward with the default observations and sparse local
observations with the standard force-coefficient reward. The observation comparison
will vary probe number, location and measured quantity using instantaneous
measurements as policy inputs. These comparisons will assess the ability
of differentiable simulation to optimise feedback policies, suppress wake
oscillations and simplify reward and observation designs in this classical
flow-control problem.

Section~\ref{sec:flow-configuration} describes the flow configuration,
jet actuation and probe measurements. Section~\ref{sec:methodology}
presents the differentiable policy optimisation method.
Section~\ref{sec:results} reports the baseline performance, feedback
control behaviour, and comparisons of reward and observation designs.
Section~\ref{sec:conclusions} summarises the findings and discusses
computational costs and directions for further assessment.

\section{Flow-control problem and numerical configuration}
\label{sec:flow-configuration}

We consider synthetic-jet control of a two-dimensional confined
circular-cylinder wake at $\Rey=100$. The flow simulation provides the
response to jet actuation, while local probes supply observations to the
control policy. The flow configuration, jet boundary conditions and
measurements are specified below; policy optimisation is described in
\S\ref{sec:methodology}.

\subsection{Flow configuration and numerical solver}
\label{sec:solver}\label{sec:computational-setup}

The flow is governed by the two-dimensional incompressible Navier--Stokes equations,
\begin{equation}\label{eq:NS-momentum}
  \frac{\partial \bfu}{\partial t}
  + (\bfu \cdot \nabla)\bfu
  = -\frac{1}{\rho}\nabla p + \nu\,\nabla^2 \bfu ,
\end{equation}
\begin{equation}\label{eq:NS-continuity}
  \nabla \cdot \bfu = 0 .
\end{equation}
Here $t$ is the non-dimensional time used by the solver,
$\bfx=(x,y)$ is the spatial position, and $\bfu(\bfx,t)=(u,v)$ comprises
the streamwise and cross-stream velocity components. Pressure, density
and kinematic viscosity are denoted by $p$, $\rho$ and $\nu$, respectively.
We use this time variable throughout, with solver increment $\Delta t$.
For the channel flow considered here, $\Rey=U_\infty D/\nu$, where
$U_\infty$ denotes the mean inlet velocity and $D$ the cylinder diameter.

The domain dimensions and Reynolds number follow the circular-cylinder
benchmark of \citet{Rabault2019}. Lengths are scaled by $D$, and the
computational domain is
\begin{equation}
  \Omega / D = [-2,20] \times [-2.05,2.05].
  \label{eq:computational-domain}
\end{equation}
The cylinder is centred between two parallel channel walls, with the
coordinate origin at its centre (figure~\ref{fig:computational-setup}).
The channel height $H$ is the distance between the two walls.
The inlet velocity has a parabolic profile with section-mean velocity
$U_\infty=2$:
\begin{equation}
  \bfu_{\mathrm{in}}(y)
  = \left[\frac{3}{2}U_\infty
      \left(1-\left(\frac{2y}{H}\right)^2\right),\,0\right]^{\mathsf{T}},
  \qquad H=4.1D .
  \label{eq:inlet-profile}
\end{equation}
No-slip conditions apply to the channel walls and to the cylinder surface
outside the jet arcs. At the outlet, a zero-gradient condition is imposed
on velocity, and pressure is fixed to its reference value. Control starts
from a developed, periodically shedding uncontrolled wake. The episode initialisation protocol is given in
\S\ref{sec:training}.

The equations are solved with $\Phi_\textrm{Flow}$
\citep{Holl2020,Holl2024PhiFlow} using its PyTorch backend
\citep{Paszke2019}. The spatial discretisation uses a staggered grid;
time stepping combines MacCormack advection with automatic diffusion
substepping. The grid resolution is $559\times104$, and the solver
time step is $\Delta t=5\times10^{-3}$.

The drag and lift coefficients are
\begin{equation}\label{eq:Cd}
  \Cd=\frac{F_x}{\tfrac12\rho U_\infty^2D},\qquad
  \Cl=\frac{F_y}{\tfrac12\rho U_\infty^2D},
\end{equation}
where $F_x$ and $F_y$ are the streamwise and cross-stream forces.

\begin{fwfigure}
  \centering
  \includegraphics[width=\linewidth]{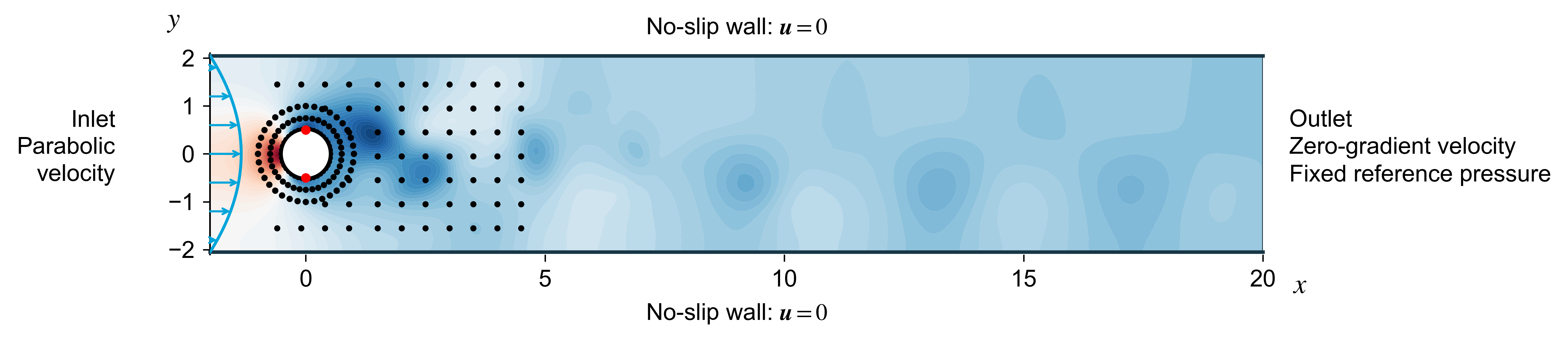}
  \caption{Computational domain for confined circular-cylinder flow control,
    showing the inlet, wall and outlet boundary conditions, the cylinder,
    fixed baseline probe layout (black dots) and jet centres (red dots).
    Coordinates are scaled by $D$ and centred on the cylinder.}
  \label{fig:computational-setup}
\end{fwfigure}

\subsection{Synthetic-jet actuation}\label{sec:jet-actuation}

Two symmetric jet arcs are centred at $\theta_1=\pi/2$ and
$\theta_2=3\pi/2$. The angle $\theta$ is measured counter-clockwise from the
positive streamwise direction, as shown in
figure~\ref{fig:jet-profile}. This convention applies to all
surface locations reported below. Each jet arc has width
$\omega=10^{\circ}=\pi/18$. The nozzle profile is cosine-like
\citep{Rabault2019,Li2022Confined}, and the differentiable jet velocity
boundary condition is
\begin{equation}
  \begin{gathered}
  \bfu_{j}(\theta,t)
  = a(t)\,g_j(\theta)\,\boldsymbol{e}_y,\\
  g_j(\theta)\propto
    \cos\!\left[\frac{\pi(\theta-\theta_j)}{\omega}\right],
  \qquad \lvert\theta-\theta_j\rvert \leq \frac{\omega}{2},
  \end{gathered}
  \label{eq:jet-profile}
\end{equation}
where $\boldsymbol{e}_y$ is the unit vector in the cross-stream ($y$)
direction, perpendicular to the incoming flow. The applied amplitude is
$a(t_n)=a_n$, and each nozzle profile $g_j$ is normalised to unit maximum.
Velocity is zero elsewhere on the cylinder. The policy output and
amplitude smoothing are defined in \S\ref{sec:policy}.

The jet flow rate per unit span is the integral of the normal velocity
over the nozzle arc $\Gamma_j$:
\begin{equation}\label{eq:jet-flow-rate}
  \begin{gathered}
  Q_j(t)=\int_{\Gamma_j}\bfu_j\cdot\boldsymbol{n}\,\mathrm{d}s
  =a(t)\int_{\Gamma_j}g_j(\theta)
       \boldsymbol{e}_y\cdot\boldsymbol{n}\,\mathrm{d}s,\\
  Q_1(t)+Q_2(t)=0 .
  \end{gathered}
\end{equation}
Here $\boldsymbol{n}$ is the unit normal pointing from the cylinder into
the fluid, and $\mathrm{d}s$ is the arc-length element. Positive $Q_j$
denotes blowing. The symmetric nozzle profiles and common applied amplitude give
equal and opposite outward flow rates, producing local blowing and
suction with zero net mass flux. Differentiation through these velocity
boundary conditions connects the applied amplitude to the flow evolution.

Jet actuation is reported using the dimensionless flow rate
\begin{equation}\label{eq:normalised-jet-flow-rate}
  q_j(t)=\frac{Q_j(t)}{U_\infty R\omega}
        =\frac{\langle u_n(t)\rangle_{\Gamma_j}}{U_\infty},
  \qquad R=\frac{D}{2},\qquad U_\infty=2 .
\end{equation}
Here $R\omega$ is the nozzle arc length and
$\langle u_n\rangle_{\Gamma_j}=Q_j/(R\omega)$ is the spatially averaged
normal velocity over the nozzle. The reference velocity is the mean
inlet velocity $U_\infty$. The jet-flow-rate histories reported below
use $q_j$.

\begin{fwfigure}[tbp]
  \centering
  \includegraphics[width=0.60\linewidth]{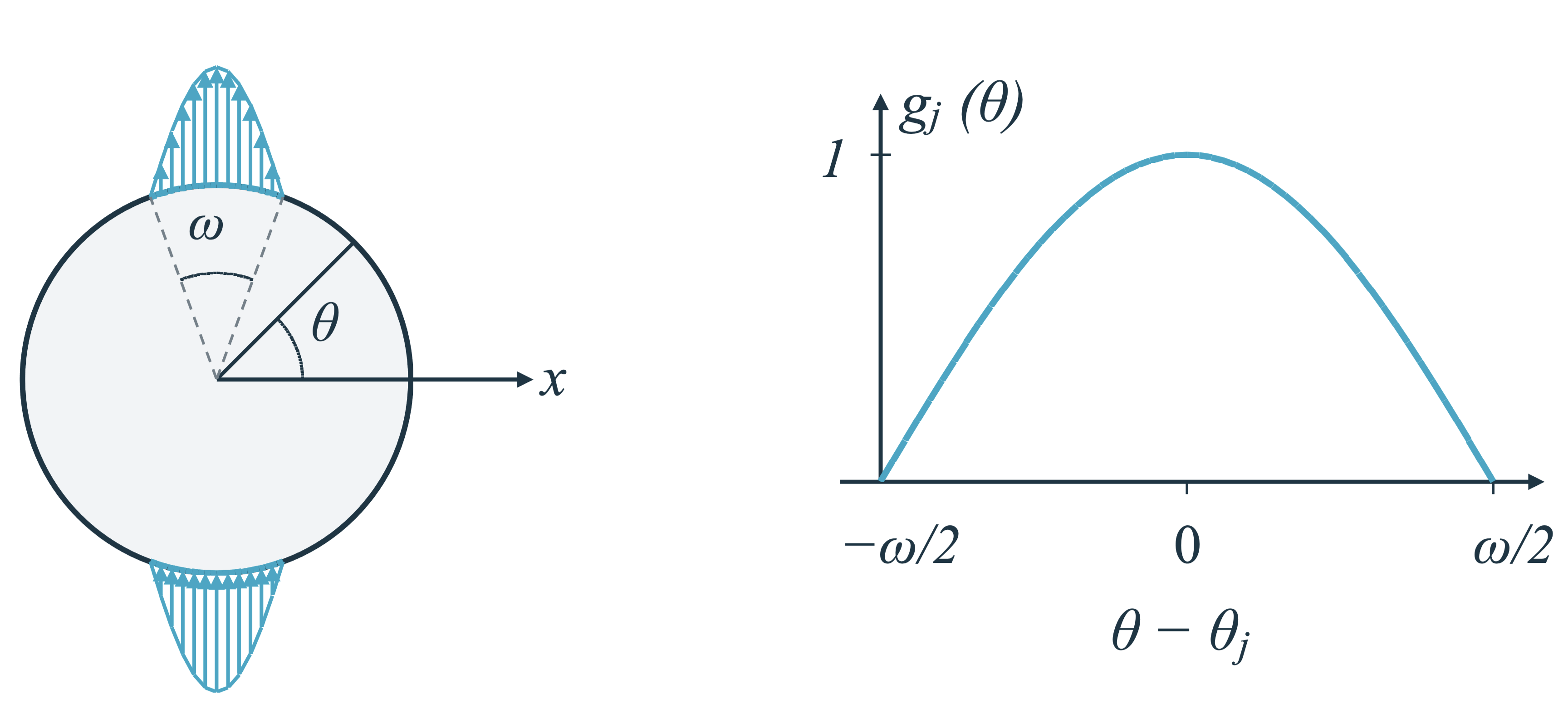}
  \caption{Synthetic-jet locations and nozzle-velocity profile.
    Left: paired jets, polar angle $\theta$ and nozzle width $\omega$;
    the arrows illustrate blowing at the upper jet and suction at the
    lower jet. Right: normalised profile $g_j(\theta)$ against the angle
    relative to the nozzle centre, $\theta-\theta_j$.}
  \label{fig:jet-profile}
\end{fwfigure}

\subsection{Flow measurements}\label{sec:measurements}

The probe observation is $\bfo=\mathcal H(\bfs)$, where $\bfs=(\bfu,p)$
is the flow state and $\mathcal H$ extracts the prescribed measurements.
The default observation comprises the two instantaneous velocity
components at 151 locations, giving an observation dimension $n_o=302$.
The probe layout follows \citet{Rabault2019} and is shown in
figure~\ref{fig:computational-setup}.

The observation comparison in \S\ref{sec:probe-configuration} also uses
single-point measurements in the near wake and on the cylinder surface
(table~\ref{tab:observation-settings}). The near-wake probe is placed
$1.5D$ downstream of the cylinder centre. Its $y$ coordinate matches that
of the surface probes at $\theta=45^{\circ}$, so the location comparison
changes only the $x$ coordinate. Surface velocity observations
(near-wall velocity observations) and surface pressure observations are
taken at the same prescribed location. All policy inputs are instantaneous
measurements, with no measurement or action history.

\begin{table}[tbp]
  \centering
  \caption{Policy-observation settings. Each velocity probe supplies
    both $(u,v)$ components; each pressure probe supplies one scalar $p$.}
  \label{tab:observation-settings}
  {\footnotesize
  \setlength{\tabcolsep}{3pt}
  \begin{tabular}{@{}>{\raggedright\arraybackslash}p{0.28\linewidth}>{\raggedright\arraybackslash}p{0.60\linewidth}c@{}}
    \toprule
    Observation & Probe locations & $n_o$ \\
    \midrule
    Default velocity array & 151 locations in figure~\ref{fig:computational-setup} & 302 \\
    Near-wake velocity & One point at $x/D=1.5$, with $y$ matched to the surface probes & 2 \\
    Surface velocity & One near-wall point at $\theta=45^{\circ}$ & 2 \\
    Surface pressure & Same location as the surface velocity probe & 1 \\
    \bottomrule
  \end{tabular}}
\end{table}

\section{Differentiable policy optimisation}\label{sec:methodology}

The method couples a neural-network policy to the flow simulation through
probe observations and differentiable synthetic-jet velocity boundary
conditions (figure~\ref{fig:framework}). In the control loop, a multilayer
perceptron (MLP) maps the current observations to a policy output, which is
limited and smoothed to obtain the applied jet amplitude.
Solver-in-the-loop policy optimisation minimises the control objective by
propagating its gradients through the simulated flow to the network
parameters using reverse-mode automatic differentiation (AD).
After optimisation, the policy parameters are fixed and the policy
controls the flow using the prescribed observations.
Training requires derivatives through the numerical flow evolution and
actuation boundary conditions over each differentiation window; deployment
uses only the trained policy and its prescribed observations.

\begin{fwfigure}[tbp]
  \centering
  \includegraphics[width=\linewidth]{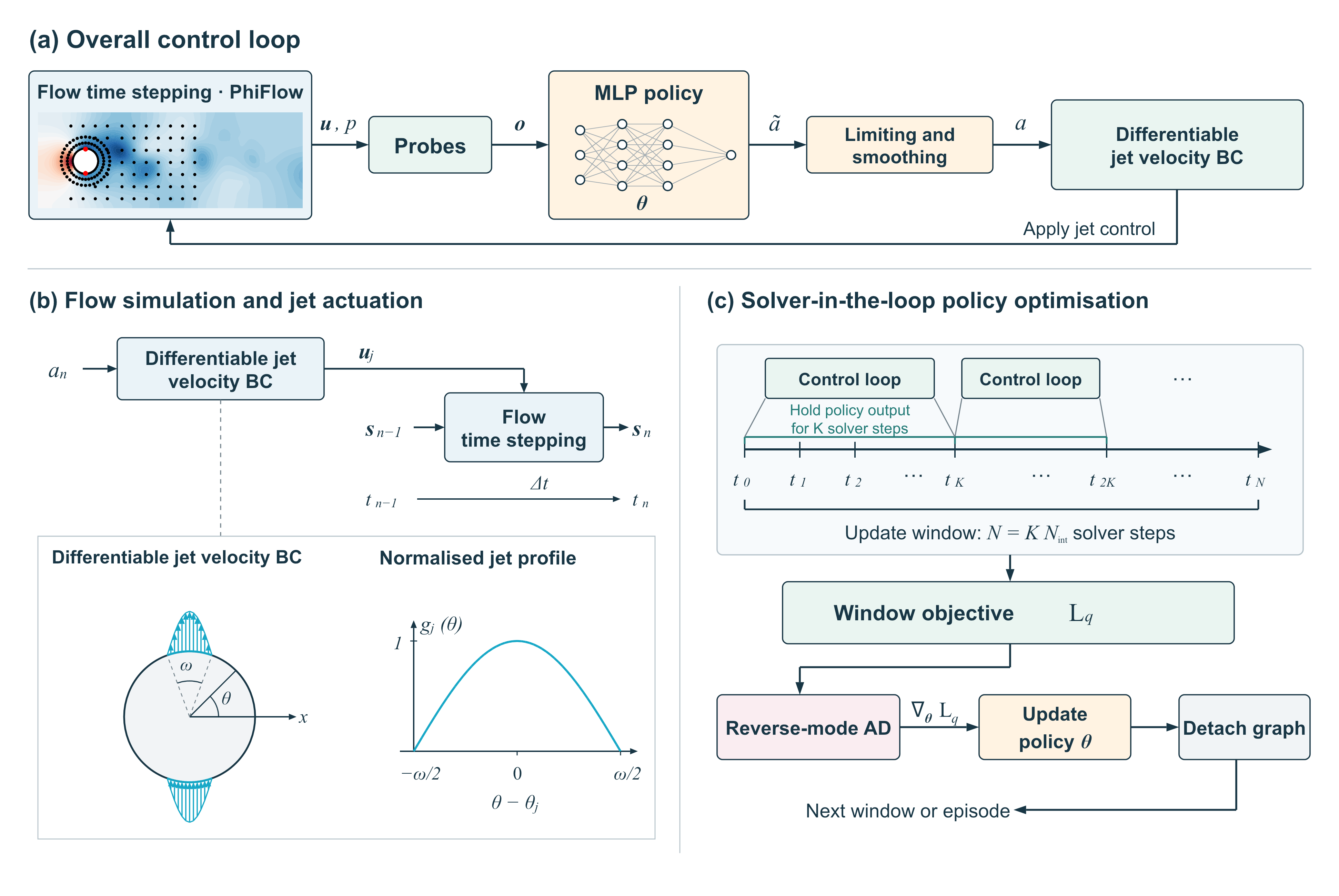}
  \caption{Differentiable active flow control framework.
  (a) Overall control loop: probe observations $\bfo$ from the flow
  state $(\bfu,p)$ enter the MLP policy with parameters
  $\boldsymbol{\theta}$. Limiting and smoothing convert its output
  $\tilde a$ into the applied jet amplitude $a$.
  Inset: flow schematic with probe and jet markers.
  (b) One solver step: the differentiable jet boundary velocity
  $\bfu_j$ and state $\bfs_{n-1}$ yield $\bfs_n$ after $\Delta t$.
  The expanded jet boundary-condition (BC) view shows the paired jets and nozzle profile
  defined in figure~\ref{fig:jet-profile}.
  (c) Solver-in-the-loop policy optimisation: an update window contains
  $N_{\mathrm{int}}$ interactions of $K$ solver steps,
  giving $N=K N_{\mathrm{int}}$. Reverse-mode AD differentiates the
  window objective $\Loss_q$ to update the policy, followed by
  detachment of the computational graph. Probe inputs are detached
  before each computation of the policy output; all other gradient
  paths remain connected within the window.}
  \label{fig:framework}
\end{fwfigure}

\subsection{Policy parameterisation and action update}
\label{sec:policy}

The policy maps the observations defined in \S\ref{sec:measurements}
to a bounded policy output. The mapping $\pi_{\boldsymbol{\theta}}:\Real^{n_o}\to\Real$
has trainable parameters $\boldsymbol{\theta}$ and observation dimension $n_o$.
It is represented by a fully connected MLP with
two hidden layers of 512 neurons each. The hidden layers use rectified
linear unit (ReLU) activations. The raw network output is mapped through
$a_{\max}\tanh(\cdot)$ to give the bounded policy output:
\begin{equation}\label{eq:policy}
  \tilde a=\pi_{\boldsymbol{\theta}}(\bfo),\qquad \tilde a\in[-a_{\max},a_{\max}].
\end{equation}
Each input component is standardised using running estimates of its
mean and variance. The normalisation statistics are fixed during
evaluation. The policy output limit is $a_{\max}=1.5$.

\paragraph{Policy interaction and flow time stepping}
Each policy interaction produces one policy output, held for $K$
solver steps while the applied jet amplitude continues to be smoothed.
Let $h$ index interactions and $r=1,\ldots,K$ the solver steps within
an interaction. The corresponding solver-step index is $n=(h-1)K+r$.
The coupled control and flow evolution is written as
\begin{equation}\label{eq:recurrence}
  \begin{aligned}
    \bfo^{(h)}&=\mathcal H(\bfs_{(h-1)K}),&
    \tilde a^{(h)}&=\pi_{\boldsymbol{\theta}}(\bfo^{(h)}),\\
    a_n&=(1-\beta)a_{n-1}+\beta c(\tilde a^{(h)}),&
    \bfs_n&=\mathcal F(\bfs_{n-1},a_n).
  \end{aligned}
\end{equation}
Here $\mathcal F$ advances the flow by one solver step,
$c(\tilde a)=\operatorname{clip}(\tilde a,-a_{\max},a_{\max})$ limits the
policy output, and $\beta=0.4$ is the temporal smoothing coefficient.
For $K=1$, a new policy output is computed at every solver step.
The training-time differentiation rules are specified in
\S\ref{sec:gradients}.

\subsection{Rewards and optimisation objective}
\label{sec:loss-design}

We consider two instantaneous losses for optimising wake suppression:
one based on drag and lift coefficients, and the other on deviations
from reference velocities at local measurement points.

Following \citet{Wang2022DRLinFluids}, the default force-coefficient
reward combines drag reduction with a penalty on lift. At solver step
$j$, it is evaluated from the instantaneous force coefficients:
\begin{equation}\label{eq:drl-reward}
  r_j=C_{d,0}-C_{d,j}-\alpha_l|C_{l,j}|,\qquad \alpha_l=0.1.
\end{equation}
The coefficient $\alpha_l$ weights the lift penalty. The baseline offset
$C_{d,0}$ is constant and contributes no gradient. Negating the reward and omitting the
constant offset gives the instantaneous loss; minimising it is equivalent to
maximising the reward:
\begin{equation}\label{eq:loss-force}
  \ell_{\mathrm{force},j}=-(r_j-C_{d,0})=C_{d,j}+0.1|C_{l,j}|.
\end{equation}
For the velocity-based reward, the instantaneous loss is the mean
squared difference between the velocity components and their reference
values at six downstream locations:
\begin{equation}\label{eq:loss-probe}
  \begin{aligned}
  \ell_{\mathrm{probe}}(t)
  &= \frac{1}{2N_{\ell}}\sum_{i=1}^{N_{\ell}}
    \left[\bigl(u(\bfx_i,t)-u_i^*\bigr)^2
          +\bigl(v(\bfx_i,t)-v_i^*\bigr)^2\right],\\
  N_{\ell} &= 6 .
  \end{aligned}
\end{equation}
The six points lie at $(x-x_c)/D=1.5,\,2.0,\,2.5$, with a pair at
$(y-y_c)/D=\pm0.25$ at each streamwise station; $(x_c,y_c)$ and $D$ denote
the cylinder centre and diameter.  These reward-evaluation measurements,
shown as squares in figure~\ref{fig:probe-loss-control}, are distinct
from the policy-observation probes.  The reference components $(u_i^*,v_i^*)$
are temporal means of saved uncontrolled-flow snapshots spanning eight
shedding periods.  At each station, the upper and lower means are
symmetrised: the $u$ components have the same magnitude and sign,
while the $v$ components have equal magnitudes and opposite signs.
Neither drag nor lift enters the reward.

\paragraph{Loss accumulation}
Within a policy-update window of $N$ controlled solver steps, let $\ell_j$
denote the instantaneous loss after step $j$. The discounted loss
$\loss_k$ accumulates these losses from step $k$ to the window end.
The window objective $\Loss_q$ averages $\loss_k$ over all starting steps:
\begin{equation}\label{eq:loss-interaction}
  \loss_k=\sum_{j=k}^{N}\gamma^{j-k}\ell_j,
\end{equation}
\begin{equation}\label{eq:loss-block}
  \Loss_q=\frac1N\sum_{k=1}^{N}\loss_k
        =\sum_{j=1}^{N}w_j\ell_j,\qquad
  w_j=\frac1N\sum_{k=1}^{j}\gamma^{j-k}.
\end{equation}
Here $q$ indexes the update window, and $\gamma$ is the temporal
discount factor \citep{Rabault2019}.  We use $\gamma=0.99$.
For $\gamma<1$, $w_j=(1-\gamma^j)/[N(1-\gamma)]$.
The weights combine the discounted contributions of each instantaneous
loss to these accumulated losses.

\subsection{Gradient calculation and truncation}
\label{sec:gradients}

In policy-gradient DRL, a likelihood-ratio estimator relates the
expected return $J$ to the policy parameters $\boldsymbol{\theta}$:
\begin{equation}\label{eq:pg}
  \nabla_{\boldsymbol{\theta}} J \;\approx\;
  \E_{\tau\sim\pi_{\boldsymbol{\theta}}}\!\Bigl[
    \sum_{n=0}^{N} \nabla_{\boldsymbol{\theta}} \log\pi_{\boldsymbol{\theta}}(a_n|\bfo_n)\,R_n
  \Bigr] ,
\end{equation}
Here $\tau$ is a sampled trajectory, $\bfo_n$ is the policy observation,
$a_n$ is the action, and $R_n$ is the return from step $n$.
When the flow evolution is differentiable, the gradient of an
instantaneous loss can instead be expressed along the path from the
loss through the flow state and actions to the policy parameters:
\begin{equation}\label{eq:diffgrad}
  \frac{\mathrm{d}\ell_n}{\mathrm{d}\boldsymbol{\theta}}
  \;=\;
  \frac{\partial\ell_n}{\partial \bfs_n}\,
  \frac{\mathrm{d} \bfs_n}{\mathrm{d} a}\,
  \frac{\mathrm{d} a}{\mathrm{d}\boldsymbol{\theta}}\,,
\end{equation}
Here $a$ collects the applied control amplitudes that influence the flow state $\bfs_n$.
The state-response factor accounts for the effects of preceding actions
propagated through the flow. Writing the one-step state evolution as
$\bfs_n=\mathcal F(\bfs_{n-1},a_n)$, the temporal expansion separates historical
contributions from the immediate action:
\begin{equation}\label{eq:bptt}
  \begin{aligned}
  \frac{\mathrm{d}\ell_N}{\mathrm{d}\boldsymbol{\theta}}
  \;=\;&
  \underbrace{
    \sum_{n=N_0+1}^{N-1}
    \frac{\partial \ell_N}{\partial \bfs_N}
    \left( \prod_{m=n+1}^{N} \frac{\partial \bfs_m}{\partial \bfs_{m-1}} \right)
    \frac{\partial \bfs_{n}}{\partial a_n}\,
    \frac{\partial a_n}{\partial \boldsymbol{\theta}}
  }_{\text{historical accumulation}}
  \\[2pt]
  &+\;
  \underbrace{
    \frac{\partial \ell_N}{\partial \bfs_N}\,
    \frac{\partial \bfs_N}{\partial a_N}\,
    \frac{\partial a_N}{\partial \boldsymbol{\theta}}
  }_{\text{immediate action}} ,
  \end{aligned}
\end{equation}
The index $N_0$ denotes the last step before the differentiation window,
and $N$ its final step. With the local indexing used in the optimisation
algorithm, $N_0=0$, so $N$ also denotes the window length.
The first term accumulates earlier actions' effects through the state
transitions; the second gives the final action's contribution.
These expressions describe the general gradient path; the stop-gradient
choices below specify its use in the present optimisation procedure.

The stop-gradient operator $\operatorname{sg}$ preserves its forward value
and truncates the computation graph. It is applied to the probe readings at
each computation of the policy output. Separately, the flow-evolution graph is detached
after each parameter update. All other gradient paths,
including the smoothing recurrence, remain connected within the window.

Truncating the graph at each update-window boundary limits the number of
time steps through which gradients propagate, reducing the risk of
exploding or vanishing gradients along long trajectories.

Under these differentiation choices, the parameter derivative of $a_n$
includes propagation through the smoothing recurrence, with the probe
inputs held fixed during differentiation. The gradient of the accumulated
objective combines the instantaneous-loss gradients with the weights $w_j$
in equation~\eqref{eq:loss-block}. Reverse-mode AD evaluates this gradient
once at the end of each update window, without a manually derived adjoint
solver. After gradient clipping and the optimiser step, the rollout graph
is detached.

\subsection{Optimisation procedure and settings}
\label{sec:training}

\begin{figure}[H]
\hrule\smallskip
{\scriptsize
\setlength{\tabbingsep}{0.4em}
\begin{tabbing}
\hspace{0.8em}\=\hspace{0.8em}\=\hspace{0.8em}\=\hspace{0.8em}\=\hspace{0.8em}\=\kill
\textbf{Input:} Flow advance $\mathcal F$, observation $\mathcal H$,\\
\> policy $\pi_{\boldsymbol{\theta}}$, loss $\ell$, policy-output limiter $c$, optimiser;\\
\> episodes $E$, warmup steps $N_{\mathrm{warm}}$,\\
\> update windows $N_{\mathrm{upd}}$ per episode,\\
\> interactions $N_{\mathrm{int}}$ per window, solver steps $K$ per interaction,\\
\> discount factor $\gamma$, smoothing coefficient $\beta$.\\[2pt]
\textbf{Set} $N\gets K N_{\mathrm{int}}$\\
\textbf{for} episode $=1$ \textbf{to} $E$ \textbf{do}\\
\> Restore the common uncontrolled state and set $a_0=0$\\
\> Advance $N_{\mathrm{warm}}$ steps without jets;\\
\> denote the resulting state by $\bfs_0$\\
\> \textbf{for} window $q=1$ \textbf{to} $N_{\mathrm{upd}}$ \textbf{do}\\
\>\> \textbf{for} interaction $h=1$ \textbf{to} $N_{\mathrm{int}}$ \textbf{do}\\
\>\>\> $\bfo^{(h)}\gets\mathcal H(\bfs_{(h-1)K})$\\
\>\>\> $\tilde a^{(h)}\gets\pi_{\boldsymbol{\theta}}(\operatorname{sg}[\bfo^{(h)}])$\\
\>\>\> \textbf{for} solver step $r=1$ \textbf{to} $K$ \textbf{do}\\
\>\>\>\> $n\gets(h-1)K+r$\\
\>\>\>\> $a_n\gets(1-\beta)a_{n-1}+\beta c(\tilde a^{(h)})$\\
\>\>\>\> $\bfs_n\gets\mathcal F(\bfs_{n-1},a_n)$; record $\ell_n=\ell(\bfs_n)$\\
\>\>\> \textbf{end for}\\
\>\> \textbf{end for}\\
\>\> $\Loss_q\gets\sum_{j=1}^{N}w_j\ell_j$, with $w_j$ from \eqref{eq:loss-block}\\
\>\> Compute $\nabla_{\boldsymbol{\theta}}\Loss_q$ by reverse-mode AD\\
\>\> Clip the global gradient norm and update $\boldsymbol{\theta}$\\
\>\> with the optimiser\\
\>\> Detach the rollout graph; set $\bfs_0\gets\bfs_N$ and $a_0\gets a_N$\\
\> \textbf{end for}\\
\textbf{end for}\\
\textbf{Output:} Optimised policy $\pi_{\boldsymbol{\theta}}$.
\end{tabbing}
}
\smallskip\hrule
\refstepcounter{algorithm}%
\caption*{Algorithm \thealgorithm: Differentiable AFC policy optimisation.}%
\label{alg:training}
\end{figure}

Let $E$ be the number of training episodes, $N_{\mathrm{warm}}$ the number
of uncontrolled warmup steps, and $N_{\mathrm{upd}}$ the number of policy-update
windows per episode. Each window contains $N_{\mathrm{int}}$ interactions
of $K$ solver steps, so $N=K N_{\mathrm{int}}$. We use $E=8$,
$N_{\mathrm{warm}}=20$, $N_{\mathrm{upd}}=1$, $N_{\mathrm{int}}=5000$ and $K=1$.
Each episode restarts from the same uncontrolled flow state, and the
network is updated at each window end. The policy after the eighth
update is used for evaluation.

The optimisation settings in table~\ref{tab:training-configuration}
apply throughout this study unless otherwise stated. Algorithm~\ref{alg:training}
combines the observation, actuation, flow-advance and gradient-update
operations into the complete policy optimisation procedure.

\begin{table}[tbp]
  \centering
  \caption{Default optimisation parameters.}
  \label{tab:training-configuration}
  {\small
  \setlength{\tabcolsep}{3pt}
  \begin{tabular}{@{}>{\raggedright\arraybackslash}p{0.60\linewidth}>{\raggedright\arraybackslash}p{0.34\linewidth}@{}}
    \hline
    Parameter & Setting \\
    \hline
    Optimiser & AdamW \\
    Base / nominal terminal learning rate & $10^{-3}$ / $10^{-5}$ \\
    Learning-rate schedule & Cosine; 2-update warmup \\
    Weight decay & $5\times10^{-4}$ \\
    Global gradient $L_2$-norm limit & $1.0$ \\
    \hline
  \end{tabular}}
\end{table}

\section{Results}\label{sec:results}

The results assess the training interactions and control performance of
differentiable policy optimisation, then examine the control behaviour
and opportunities to simplify its design. Section~\ref{sec:benchmark-validation}
compares control performance and cumulative training interactions
with published DRL results. Section~\ref{sec:control-behaviour} examines persistent suppression,
starting-phase dependence and disturbance recovery using the same policy.
Section~\ref{sec:design-implications} then considers the reward and the
observations in separate comparisons.

\subsection{Baseline wake and control performance}
\label{sec:benchmark-validation}\label{sec:control-performance}

\begin{fwfigure}[tbp]
  \centering
  \begin{minipage}[t]{0.49\linewidth}
    \raggedright\textbf{(a)}\par
    \includegraphics[width=\linewidth]{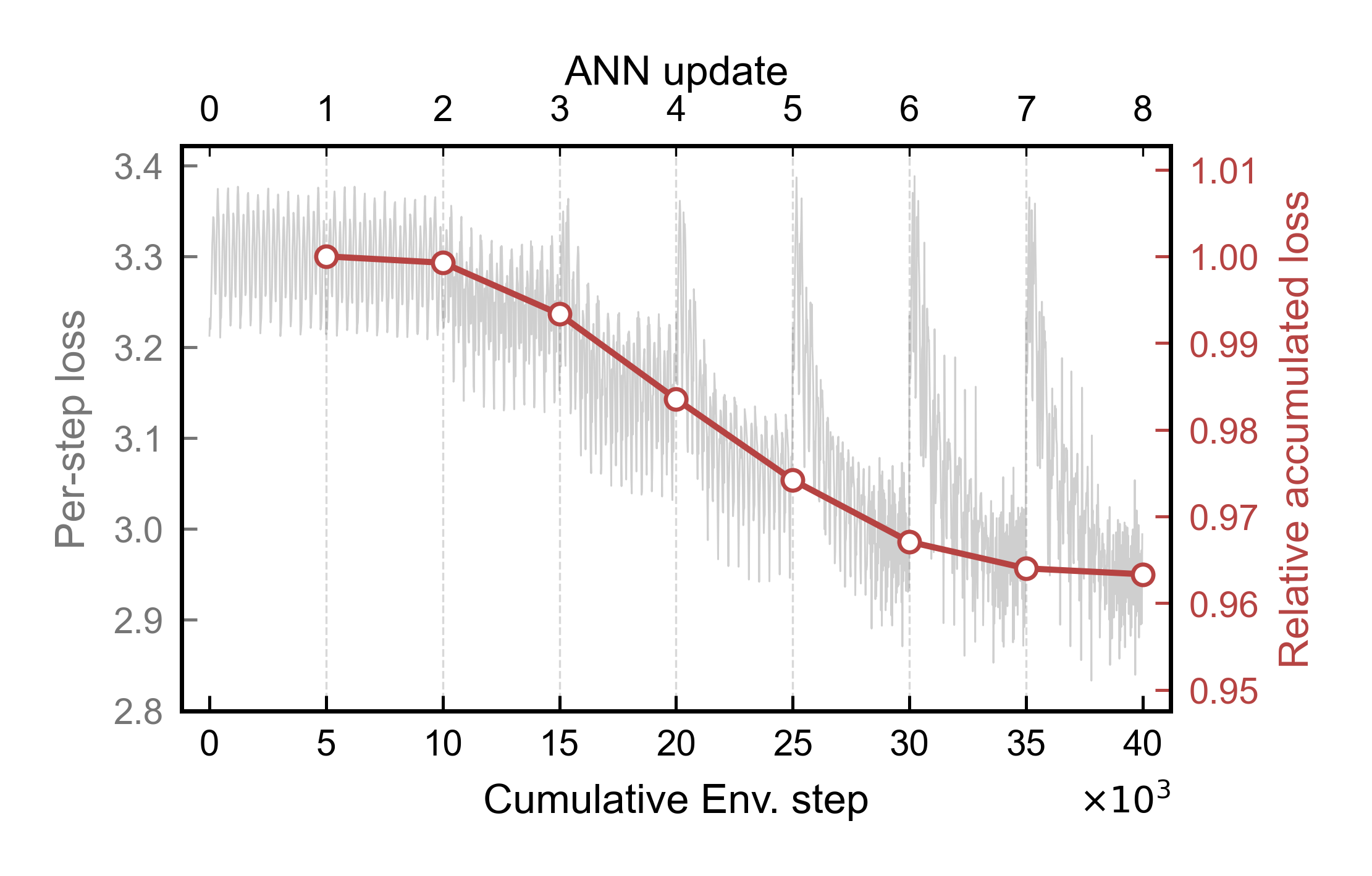}
  \end{minipage}\hfill
  \begin{minipage}[t]{0.49\linewidth}
    \raggedright\textbf{(b)}\par
    \vspace{14pt}
    \centering
    \includegraphics[width=\linewidth]{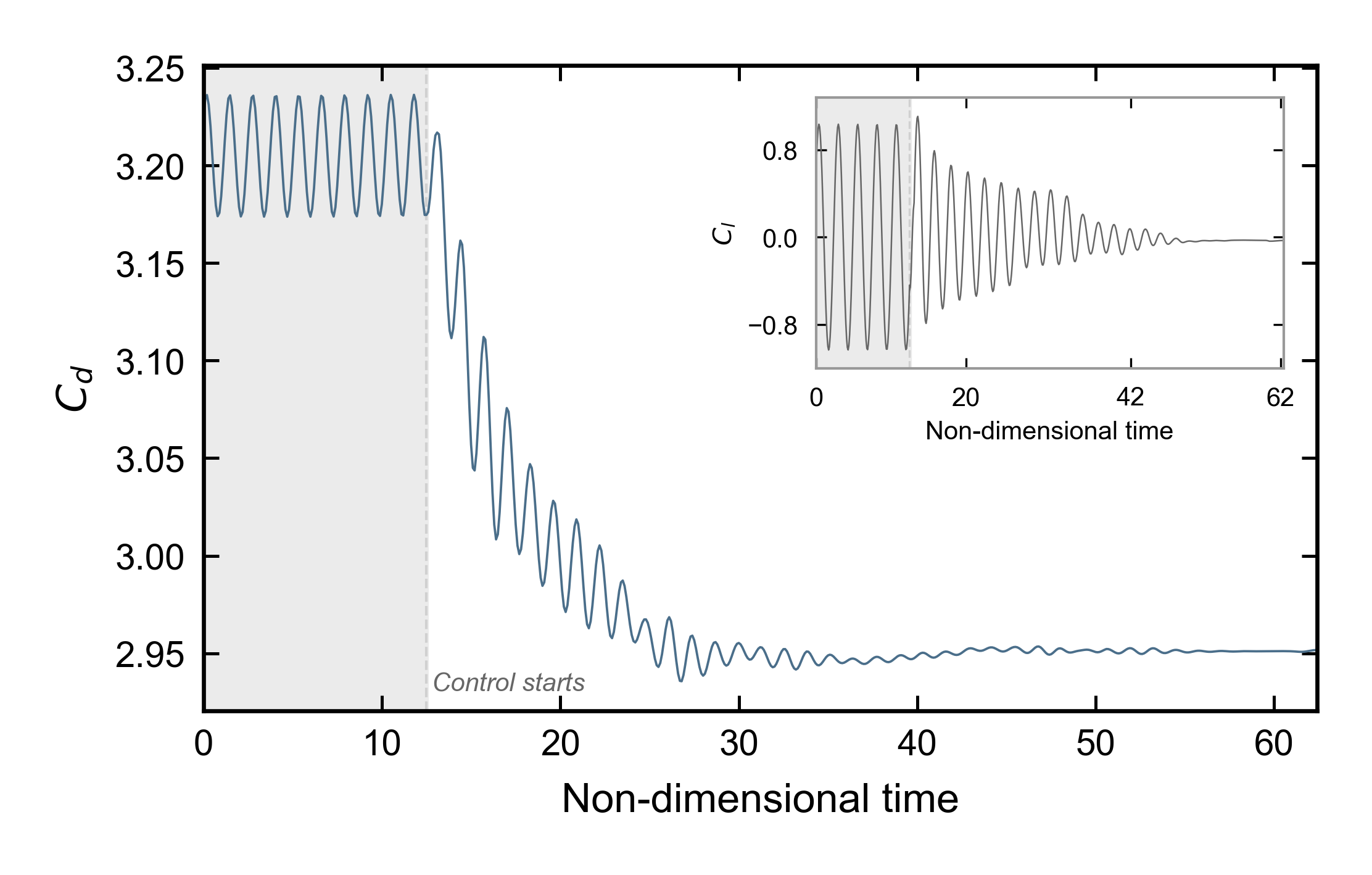}
  \end{minipage}
  \par\medskip
  \begin{minipage}{0.68\linewidth}
    \raggedright\textbf{(c)}\par
    \centering
    \includegraphics[width=\linewidth]{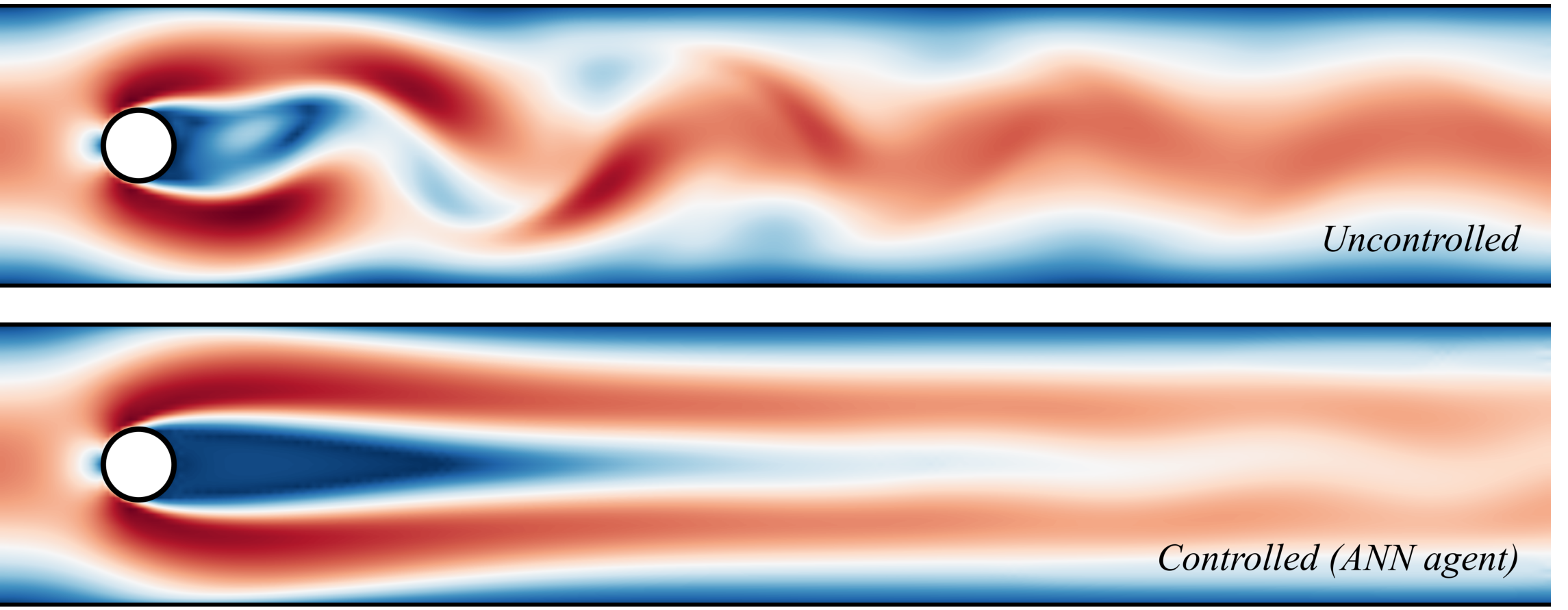}
  \end{minipage}
  \caption{Policy optimisation and baseline control performance with 151 velocity probes.
    (a) Optimisation history: the grey trace is the per-step loss $\ell_j$,
    and the red markers show the accumulated loss normalised by its
    first-episode value. The lower axis counts cumulative environment
    steps (one controlled solver step per environment step for $K=1$);
    the upper axis counts policy updates.
    (b) Drag-coefficient history, with the lift-coefficient history in the
    inset. The grey interval precedes control activation.
    (c) Uncontrolled (top) and controlled (bottom) wake structures,
    shown through velocity-magnitude fields.}
  \label{fig:control-performance}
\end{fwfigure}

The uncontrolled wake provides the reference for evaluating the policy.
The force-coefficient means and standard deviations are calculated over
the final five uncontrolled shedding periods of each evaluation record. The uncontrolled baseline yielded
$\overline{\Cd}\approx3.205$, consistent with the uncontrolled value reported
by \citet{Rabault2019}. With the 151-probe observations defined in
\S\ref{sec:measurements}, we compare the optimised policy's drag reduction,
lift-fluctuation suppression and cumulative training interactions with
published DRL results.

Over eight policy updates, the accumulated loss, normalised by its
first-episode value, decreases from 1 to approximately 0.963
(figure~\ref{fig:control-performance}(a)).  The per-step loss follows the same
decreasing envelope, with only slight changes in the final two updates.
The 40,000 cumulative controlled solver steps correspond to approximately $78$
vortex-shedding periods of the uncontrolled wake, within which the loss
progressively approaches a stable value.

After actuation begins at $t\approx12.5$, the $C_d$ envelope decreases
from an uncontrolled mean of $3.2052$ towards a late-time mean of $2.9510$
(figure~\ref{fig:control-performance}(b)).  Over the same transition, lift
fluctuations contract to a near-zero level (inset), and the standard
deviation of $C_l$ decreases from $0.7390$ to $0.0029$.  The controlled mean and
standard deviation refer to the late-time regime after the suppression
transient.

The wake fields reflect this change in the force histories.  The
uncontrolled near wake has an alternating structure, whereas the controlled
snapshot is nearly symmetric, with an elongated low-speed region downstream
of the cylinder (figure~\ref{fig:control-performance}(c)).  These features
are consistent with the simultaneous reductions in drag and lift fluctuations.

The drag-reduction and lift-fluctuation-suppression rates are defined by
\begin{align}
  \Delta\Cd(\%) &= 100\%\,
  \frac{\overline{C_d}^{\mathrm{u}}-\overline{C_d}^{\mathrm{c}}}
       {\overline{C_d}^{\mathrm{u}}}, \\
  \Delta\sigma(\Cl)(\%) &= 100\%\,
  \frac{\sigma(\Cl)^{\mathrm{u}}-\sigma(\Cl)^{\mathrm{c}}}
       {\sigma(\Cl)^{\mathrm{u}}}.
\end{align}
Here, the overbar denotes a time mean and $\sigma$ a standard deviation;
superscripts $\mathrm{u}$ and $\mathrm{c}$ denote the uncontrolled
and controlled cases, respectively.  With zero-net-mass-flux actuation
through the two jets, the policy achieves approximately $8\%$ drag
reduction and $99.6\%$ lift-fluctuation suppression.  Its mean drag,
$\overline{C_d}=2.9510$, is close to the no-shedding reference
$C_{d\mathrm{s}}=2.93$ reported by \citet{Rabault2019}.

The training budget, $\mathcal{B}$, is expressed in vortex-shedding periods of
the uncontrolled wake,
\begin{equation}
  \mathcal{B} = \frac{N_{\mathrm{step}}\,\Delta t}{T_{\mathrm{vs}}},
  \label{eq:training-budget}
\end{equation}
where $N_{\mathrm{step}}$ counts controlled solver steps across all
training environments, and $T_{\mathrm{vs}}$ is the uncontrolled shedding
period in the same time units as $\Delta t$.  The present budget comprises
eight episodes reset to the same initial state.  With
$T_{\mathrm{vs}}\approx2.56$, it gives $40{,}000\times0.005=200$ and
$200/2.56\approx78$.

Table~\ref{tab:drl-comparison} compares control performance and cumulative
training interactions with two published PPO studies. Their budgets are estimated
at the late training stages reported in each study:
approximately $2.3\times10^3$ shedding periods over about 350 episodes
for \citet{Rabault2019}, and $2.6\times10^2$ periods over about 350 episodes
per environment with four parallel environments for \citet{Wang2022DRLinFluids}.
The latter conversion uses the shedding period read from the published
uncontrolled time history. Both studies report approximately $8\%$
drag reduction, with lift-fluctuation suppression of approximately
$82.5\%$ and $95.68\%$, respectively, compared with $99.6\%$ in the present
study using about $78$ shedding periods of training.

\begin{fwtable}
  \centering
  \caption{Comparison with published DRL results for circular-cylinder control.
    $N_p$ is the number of velocity probes.  The cumulative budget
    $\mathcal{B}$ in \eqref{eq:training-budget} sums all training
    environments.}
  \label{tab:drl-comparison}
  {\setlength{\tabcolsep}{2pt}
  \renewcommand{\arraystretch}{1.0}
  \scriptsize
  \begin{tabular}{@{}p{0.17\textwidth}p{0.24\textwidth}cccc@{}}
    \hline
    Study & Method & $N_p$ & $\Delta\Cd\,(\%)$ &
    $\Delta\sigma(\Cl)\,(\%)$ & Training budget \\
    &&& (drag reduction) & (lift suppression) & (periods) \\
    \hline
    Ours & Differentiable-simulation ANN & 151 & $\sim8$ & $99.6$ & $\sim78$ \\
    \citet{Rabault2019} & DRL (PPO) & 151 & $\sim8$ & $\sim82.5$ & $\sim2.3\times10^3$ \\
    \citet{Wang2022DRLinFluids} & DRL (PPO) & 149 & $\sim8$ & $95.68$ & $\sim2.6\times10^2$ \\
    \hline
  \end{tabular}
  }
\end{fwtable}

The policy thus combines drag reduction with strong lift-fluctuation
suppression at a lower interaction budget than the two published DRL
studies. These force and flow-field results establish the reference control
response; the next comparisons examine how the policy maintains
suppression and responds to changes in the flow.

\subsection{Feedback control behaviour}
\label{sec:control-behaviour}

With the baseline performance established, we examine the behaviour of the
policy optimised with the conventional force-coefficient reward. The
following tests address three aspects of feedback control: maintenance of
the suppressed wake, response to different shedding phases at activation,
and recovery after upstream disturbances. The policy and observation
layout remain fixed across these tests.

\subsubsection{Complete suppression and persistent oscillatory control}
\label{sec:complete-and-persistent}

We first examine whether the policy optimised with the force-coefficient reward
suppresses shedding and maintains the non-shedding wake. In \citet{Li2022Confined},
real-time jet adjustments maintained a wake close to a potentially linearly
unstable base flow obtained by selective frequency damping (SFD). The remaining jet oscillations were essential to
continuous suppression. Here, we examine the same physical question using
the conventional force-coefficient reward, without a base-flow reference or
instability penalty in the reward.

\begin{figure}[tbp]
  \centering
  \includegraphics[width=\linewidth]{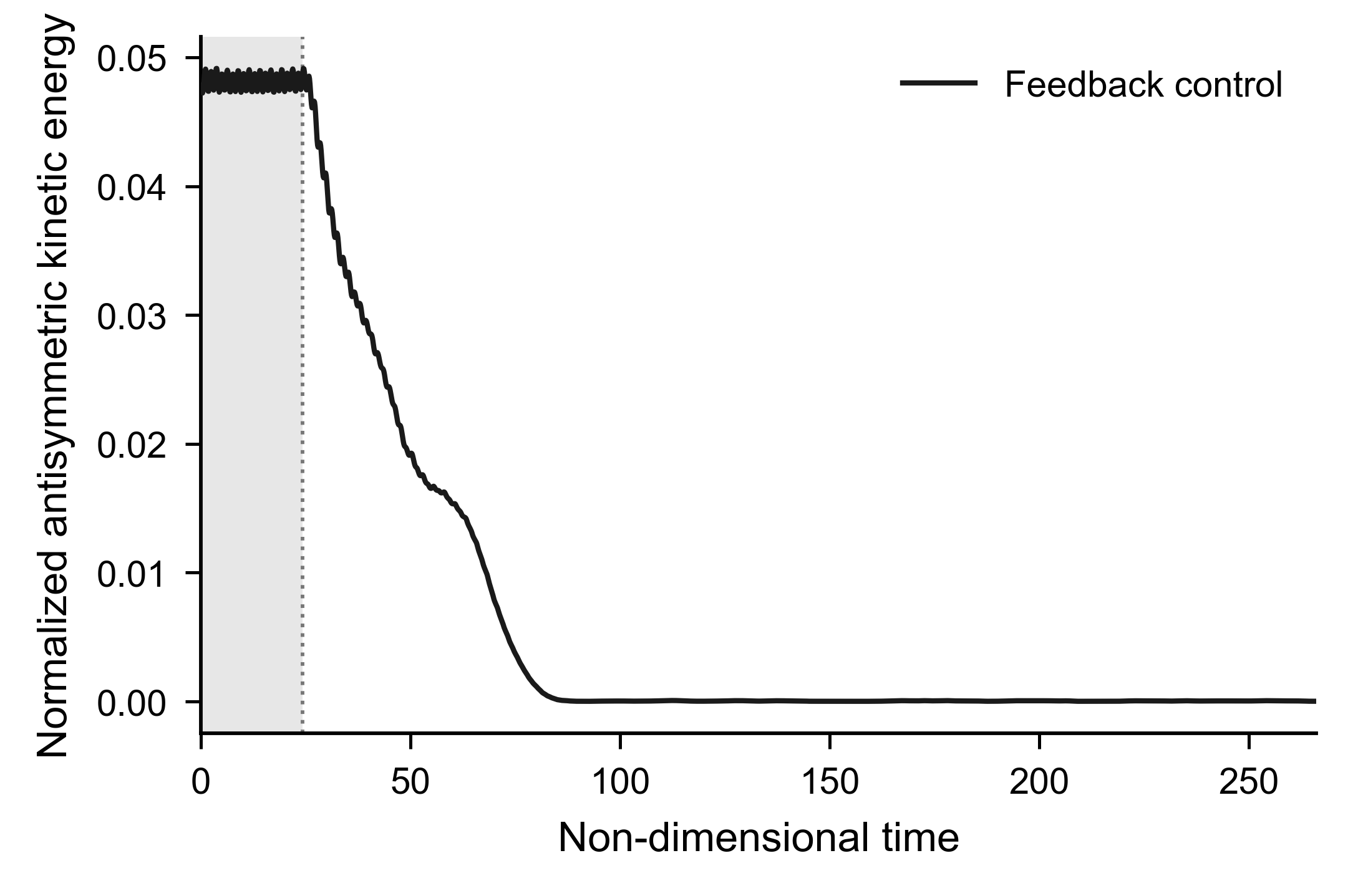}
  \caption{Decay of the normalized antisymmetric kinetic energy under
    feedback control.  The grey interval precedes feedback
    control, and the vertical grey dotted line marks its onset.}
  \label{fig:complete-suppression-energy}
\end{figure}

\begin{fwfigure}[tbp]
  \centering
  \includegraphics[width=\linewidth]{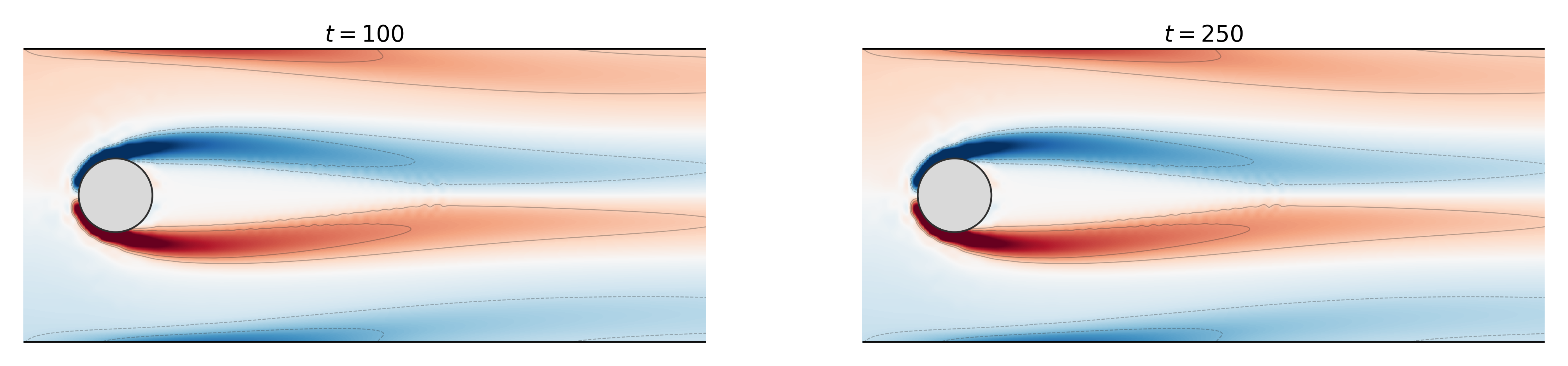}
  \caption{Spatial distribution of controlled wake vorticity at two late-time instants:
    $t=100$ (left) and $t=250$ (right).  Red and blue distinguish
    opposite signs of the spanwise vorticity.  Only part of the
    computational domain is shown.}
  \label{fig:controlled-vorticity}
\end{fwfigure}

We assess suppression using the centreline-antisymmetric kinetic energy,
lift fluctuations and late-time vorticity fields.  To test the maintenance
of the non-shedding state, we then replace the late-time feedback signal
with mean-matched constant jet actuation and compare the lift response and
dynamic mode decomposition (DMD) diagnostics.  The centreline mirror operator
and normalized antisymmetric kinetic energy are defined as
\begin{equation}
  \begin{aligned}
    \mathcal{M}\bfu(x,y,t)
      &= \begin{bmatrix}u(x,-y,t)&-v(x,-y,t)\end{bmatrix}^{\mathsf T}, \\
    \bfu_A(x,y,t)
      &= \frac{1}{2}\left[\bfu(x,y,t)-\mathcal{M}\bfu(x,y,t)\right], \\
    E_A^*(t)
      &= \frac{\sum_{\bfx_i\in\Omega_f}
          \lVert\bfu_A(\bfx_i,t)\rVert_2^2}
         {N_f U_\infty^2},
         \qquad N_f=\lvert\Omega_f\rvert .
  \end{aligned}
  \label{eq:normalized-antisymmetric-kinetic-energy}
\end{equation}

where $\Omega_f=\{\bfx_i\}_{i=1}^{N_f}$ contains the fluid sampling points
throughout the computational domain.  An exactly symmetric flow has
$E_A^*=0$, while a finite value measures symmetry breaking
\citep{desBoscsKuhlmann2024}.

Before feedback control, $E_A^*$ oscillates around
$4.8\times10^{-2}$ (figure~\ref{fig:complete-suppression-energy}).  It then
falls below $10^{-2}$ by $t\approx70$ and approaches zero by
$t\approx85$.  This near-zero level is maintained, without a
subsequent recovery of the symmetry-breaking component, through the end of
the record at $t\approx265$.


At $t=100$ and $250$, the wake remains approximately centreline symmetric,
without the alternating roll-up characteristic of vortex shedding
(figure~\ref{fig:controlled-vorticity}).  Together with the sustained
near-zero $E_A^*$ and strongly suppressed lift fluctuations in
figure~\ref{fig:control-performance}, these widely separated snapshots
demonstrate complete and persistent suppression over the observed interval.
The symmetric topology is also consistent with the SFD-base-flow signature
reported for completely suppressed confined-cylinder wakes
\citep{Li2022Confined}.

\begin{fwfigure}[tbp]
  \centering
  \includegraphics[width=\linewidth]{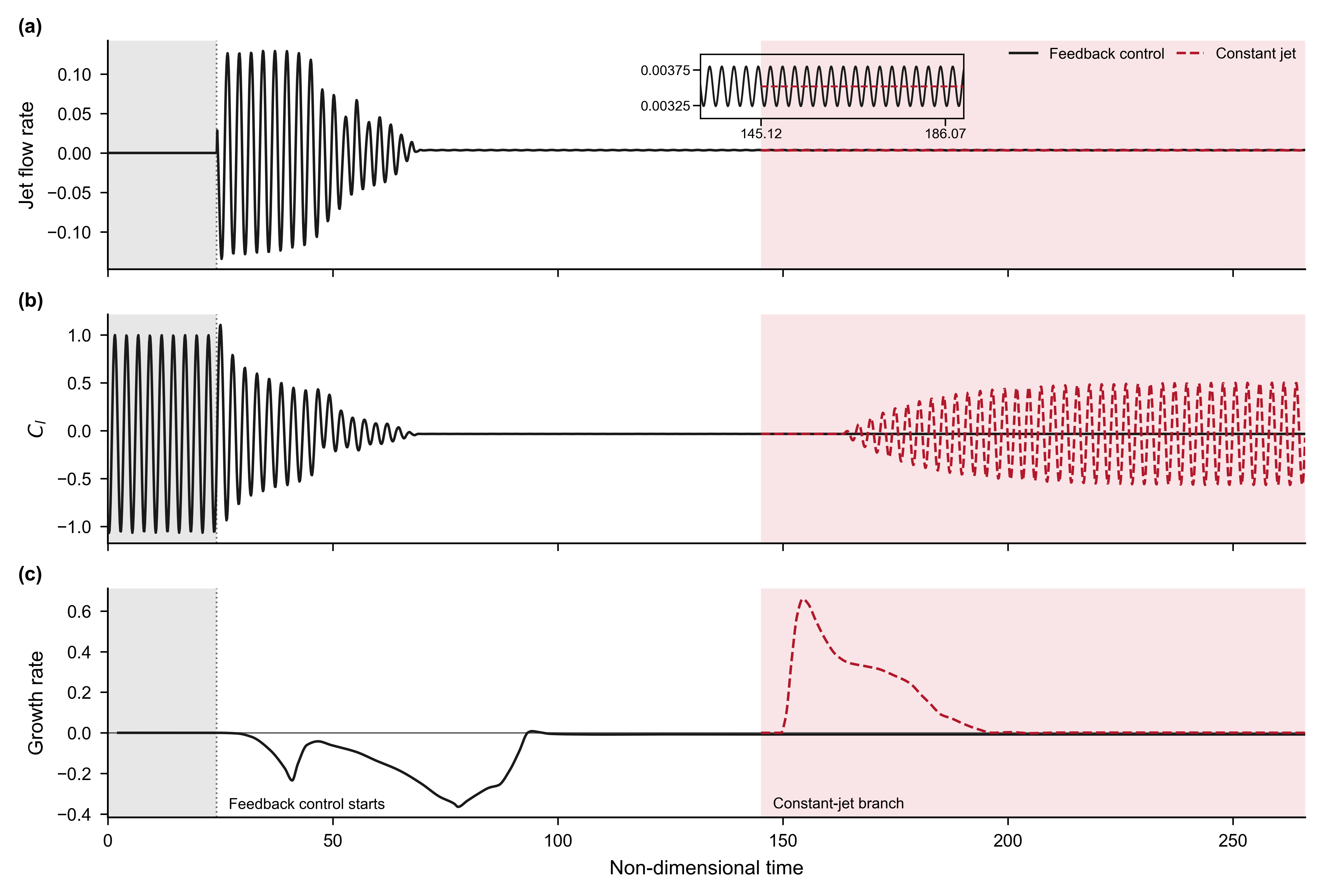}
  \caption{Temporal evolution under feedback control and in the
    constant-jet branch.  (a) Jet flow rate $q_j$ defined in
    equation~\eqref{eq:normalised-jet-flow-rate}; the inset
    expands the late-time feedback oscillation and the constant jet flow rate.
    (b) Lift coefficient $\Cl$.  (c) Growth rate of the leading DMD mode.
    The grey interval precedes feedback control; the red-shaded interval
    begins when the constant-jet branch is introduced.  Black solid and red
    dashed lines denote feedback control and the constant-jet branch,
    respectively.}
  \label{fig:persistent-control-dmd}
\end{fwfigure}

To determine whether the small feedback modulation sustains this state,
we branch the controlled trajectory at $t=145.12$.  One branch retains
feedback; the other replaces the late-time oscillatory actuation with a
constant dimensionless jet flow rate $q_j\approx3.5\times10^{-3}$, close to the
feedback mean (figure~\ref{fig:persistent-control-dmd}(a)).  This mean-matched
intervention removes temporal modulation while preserving the mean
actuation level.  DMD characterises the local evolution of flow-field
snapshots \citep{Li2022Confined}, with the leading mode defined as that with
the largest estimated growth rate (Appendix~\ref{app:dmd}).

At the onset of feedback control, the jet flow rate $q_j$ reaches an oscillation
amplitude of approximately $0.13$ as the $\Cl$ oscillation decays from its
uncontrolled level (figure~\ref{fig:persistent-control-dmd}(a,b)). Once the
non-shedding state is established, the jet flow rate oscillates between
approximately $0.00325$ and $0.0038$, while $\Cl$ remains close to zero.
Over the same transition, the leading-mode growth rate becomes negative and
then approaches zero under continued feedback
(figure~\ref{fig:persistent-control-dmd}(c)). Near-zero lift fluctuations
coexist with small but finite jet-flow-rate oscillations, consistent with
the active-maintenance mechanism identified in \citet{Li2022Confined}.

\begin{fwfigure}[tbp]
  \centering
  \includegraphics[width=\linewidth]{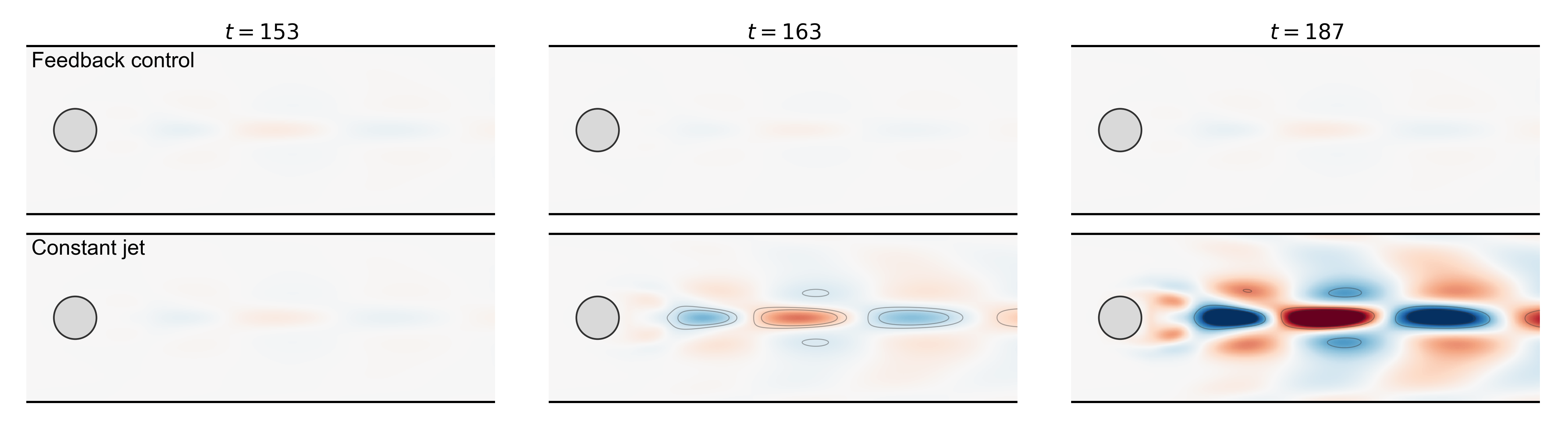}
  \caption{Vorticity structure of the leading DMD mode under feedback
    control (top row) and constant-jet forcing (bottom row) at
    $t=153$, $163$ and $187$ (left to right).  Red and blue distinguish
    opposite signs of the modal vorticity.}
  \label{fig:leading-dmd-modes}
\end{fwfigure}

With this modulation removed, the leading-mode growth rate becomes
positive and reaches approximately $0.65$.  The $\Cl$ oscillation re-emerges
from about $t=160$ and grows to a late-time amplitude of order $0.5$
(figure~\ref{fig:persistent-control-dmd}(b,c)).  Continued feedback instead
maintains a near-zero lift response and prevents sustained positive modal
growth.

The leading-mode structures are similar in the two branches at $t=153$,
but diverge by $t=163$ and $187$.  The constant-jet branch develops a clear
alternating pattern downstream of the cylinder, whereas the feedback mode
remains nearly unchanged, without a comparable vortex-street signature
(figure~\ref{fig:leading-dmd-modes}).  This spatial change corroborates the
positive modal growth and recovery of finite-amplitude lift oscillations
in figure~\ref{fig:persistent-control-dmd}. In this mean-matched comparison,
persistent low-amplitude feedback sustains vortex-shedding suppression,
whereas constant actuation at the same mean jet flow rate does not.

\begin{fwfigure}[tbp]
  \centering
  \includegraphics[width=\linewidth]{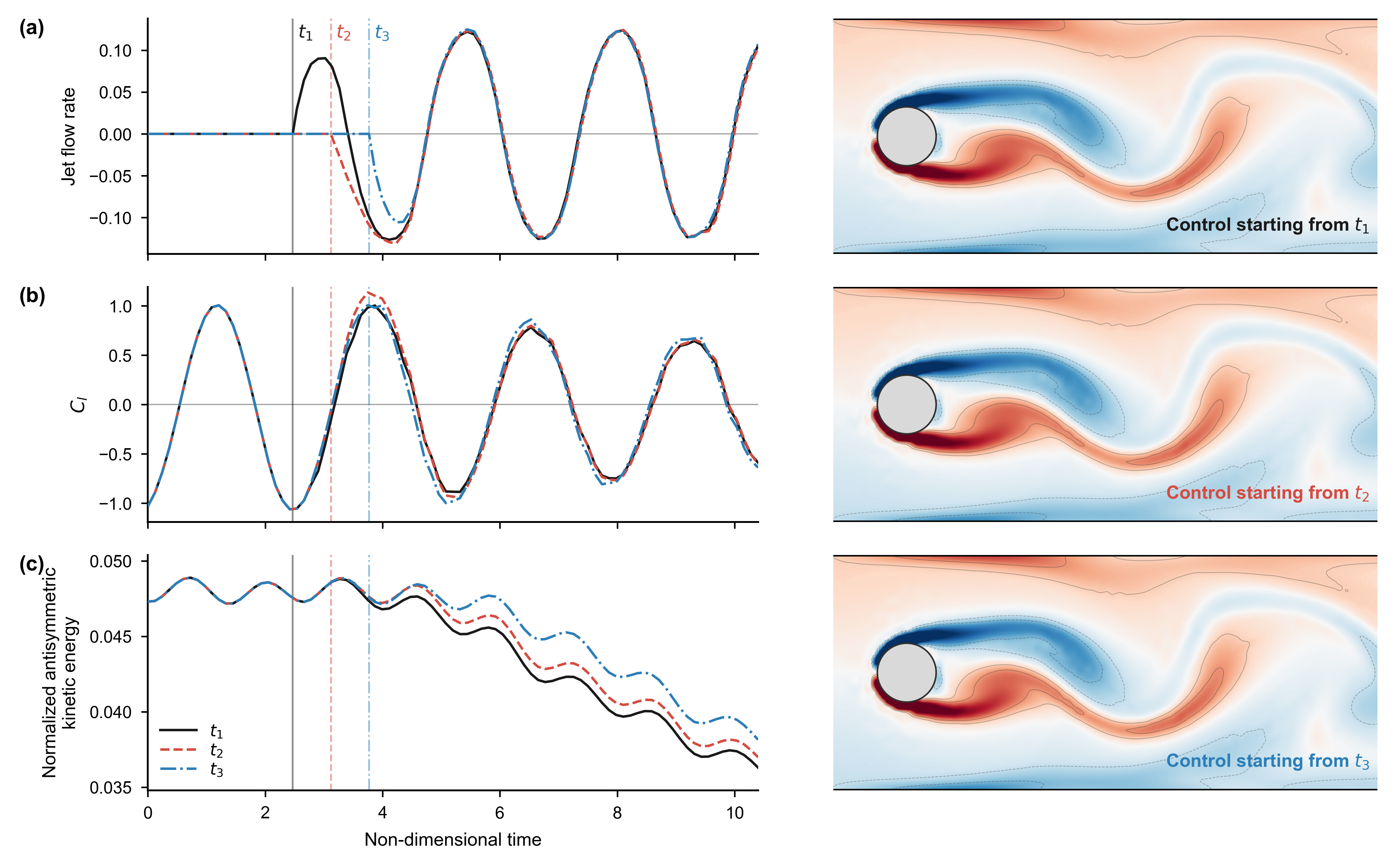}
  \caption{Effect of the control starting time with the same optimised policy
    and observation layout.
    The left column shows the jet flow rate $q_j$, lift coefficient $\Cl$
    and normalized antisymmetric kinetic energy for control activated at
    $t_1$, $t_2$ and $t_3$; the corresponding vertical lines mark the three
    activation times. The right column compares wake structures for the three
    starting phases through vorticity fields at the first post-activation
    minimum of $\Cl$ in each case.}
  \label{fig:control-starting-time}
\end{fwfigure}

\subsubsection{Control starting time}
\label{sec:starting-time}

We activate the policy at three starting times, $t_1$, $t_2$ and $t_3$,
corresponding to different phases of the uncontrolled lift oscillation. The
policy optimised with the force-coefficient reward and its observation
configuration remain fixed, isolating the response to the flow state at
activation. Figure~\ref{fig:control-starting-time} compares the jet flow
rate, lift coefficient and normalized antisymmetric kinetic energy defined
in equation~\eqref{eq:normalized-antisymmetric-kinetic-energy}.

The starting phase changes the initial actuation: the jet flow rate first
becomes positive and then changes sign for $t_1$, whereas it is initially
negative for $t_2$ and $t_3$. The antisymmetric kinetic energy decreases
in all three cases. Following the initial actuation, the three jet histories nearly
coincide, and the lift oscillations have similar phases and amplitudes,
with amplitudes decreasing over subsequent cycles.

At the first post-activation minimum of $\Cl$ in each case, the shear
layers and vortices immediately downstream of the cylinder have similar
shapes (figure~\ref{fig:control-starting-time}, right column). Together
with the jet and lift histories, these fields show that the same optimised
policy produces similar subsequent control responses across the three
tested starting phases, with a consistent decay of lift oscillations and
wake asymmetry.

\begin{fwfigure}[tbp]
  \centering
  \includegraphics[width=\linewidth]{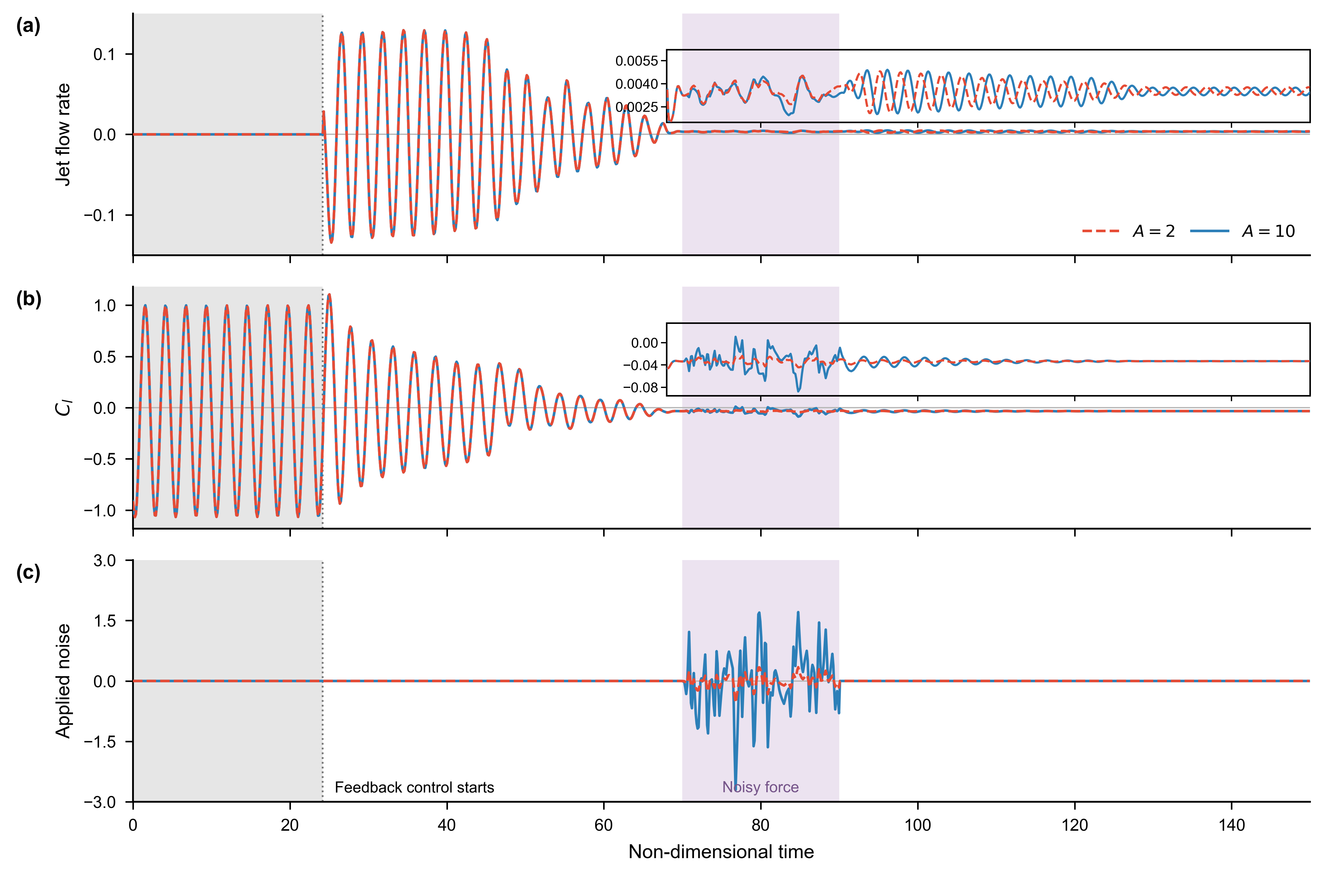}
  \caption{Response of the feedback-controlled wake to upstream noise.
    (a) Upper-jet flow rate $q_1$; the inset
    expands the low-amplitude controlled interval.  (b) Lift coefficient
    $\Cl$; the inset expands its controlled response.  (c) Applied noisy-force
    amplitude $A\xi(t)$.  The grey interval precedes feedback control, and
    the vertical grey dotted line marks its onset at $t=24$.  Purple shading
    marks the forcing interval $70\leq t\leq90$.  Red dashed and blue solid
    lines denote $A=2$ and $10$, respectively.}
  \label{fig:upstream-noise-robustness}
\end{fwfigure}

\subsubsection{Robustness to upstream noise}
\label{sec:robustness-noise}

We test disturbance rejection after the policy has established a
low-lift-fluctuation state. The policy optimised with the
force-coefficient reward and its observation configuration remain fixed. A
spatially localised upstream force \citep{Li2022Confined} with amplitude
factor $A\xi(t)$ is
applied over $70\leq t\leq90$, with $A=2$ and $10$.
Appendix~\ref{app:upstream-disturbance} specifies its spatial profile and random
scaling factor. Figure~\ref{fig:upstream-noise-robustness} shows the
upper-jet flow rate, lift coefficient and applied force before, during and
after the disturbance.

Before forcing, both cases follow the same closed-loop trajectory: the
large start-up actuation decays to a small oscillation as lift fluctuations
decrease (figure~\ref{fig:upstream-noise-robustness}(a,b)).  During forcing,
$A=10$ produces larger and more irregular $\Cl$ excursions than $A=2$.
The enlarged traces show continued variation of the jet flow rate, with
distinct actuation histories for the two disturbance amplitudes.

After the force is removed at $t=90$, the irregular lift excursions give
way to decaying oscillations.  The jet flow rate continues to oscillate,
with phase differences between the two cases, before its amplitude
decreases towards the end of the record at $t=150$.  Both lift histories
approach the same low-fluctuation level. The policy thus restores
suppression after either finite-duration upstream disturbance, without
retraining.

These comparisons show that the policy maintains suppression through
small, time-varying jet adjustments and accommodates the tested changes in
starting phase and upstream forcing. The conventional force-coefficient
reward is sufficient to learn these responses without additional
stability-based reward terms. We next examine whether simpler rewards
and observations can retain effective control.

\subsection{Effects of reward and observation design}
\label{sec:design-implications}

The reward specifies the control objective during optimisation, whereas
the observations supply feedback to the deployed policy. We examine these
two information requirements separately: the reward comparison retains the
default velocity-probe array, and the observation comparison retains the
force-coefficient reward.

\subsubsection{Simplification of the reward function}
\label{sec:loss-simplification}

Reward design can affect the quality of learned policies.
Unsuitable reward shaping can lead to policies that are suboptimal for the
original task \citep{Ng1999Policy}, while inappropriate reward scaling can
prevent effective policy learning in deep deterministic policy gradient
experiments \citep{Henderson2018Deep}. For confined-cylinder wakes,
\citet{Li2022Confined} demonstrated how physical information can guide
reward design for DRL control. A shedding-energy reward referenced to an
SFD base flow enabled complete suppression, and a DMD growth-rate term
improved persistence at higher Reynolds number. Here we ask whether direct
flow-response gradients can support effective suppression with a simpler
reward based on local velocity deviations from a symmetrised time-mean
flow, without base-flow calculations or instability penalties.

The velocity-based reward is tested with a separately optimised policy,
retaining the default 151-probe observations and shared settings in
\S\ref{sec:training}. Comparison with the force-coefficient
reward addresses both shedding suppression and the suppression transient.

The velocity-based loss in equation~\eqref{eq:loss-probe} uses the six
reward-evaluation sites shown in figure~\ref{fig:probe-loss-control}; these
are distinct from the policy-observation probes.

\begin{fwfigure}[tbp]
  \centering
  \includegraphics[width=0.80\linewidth]{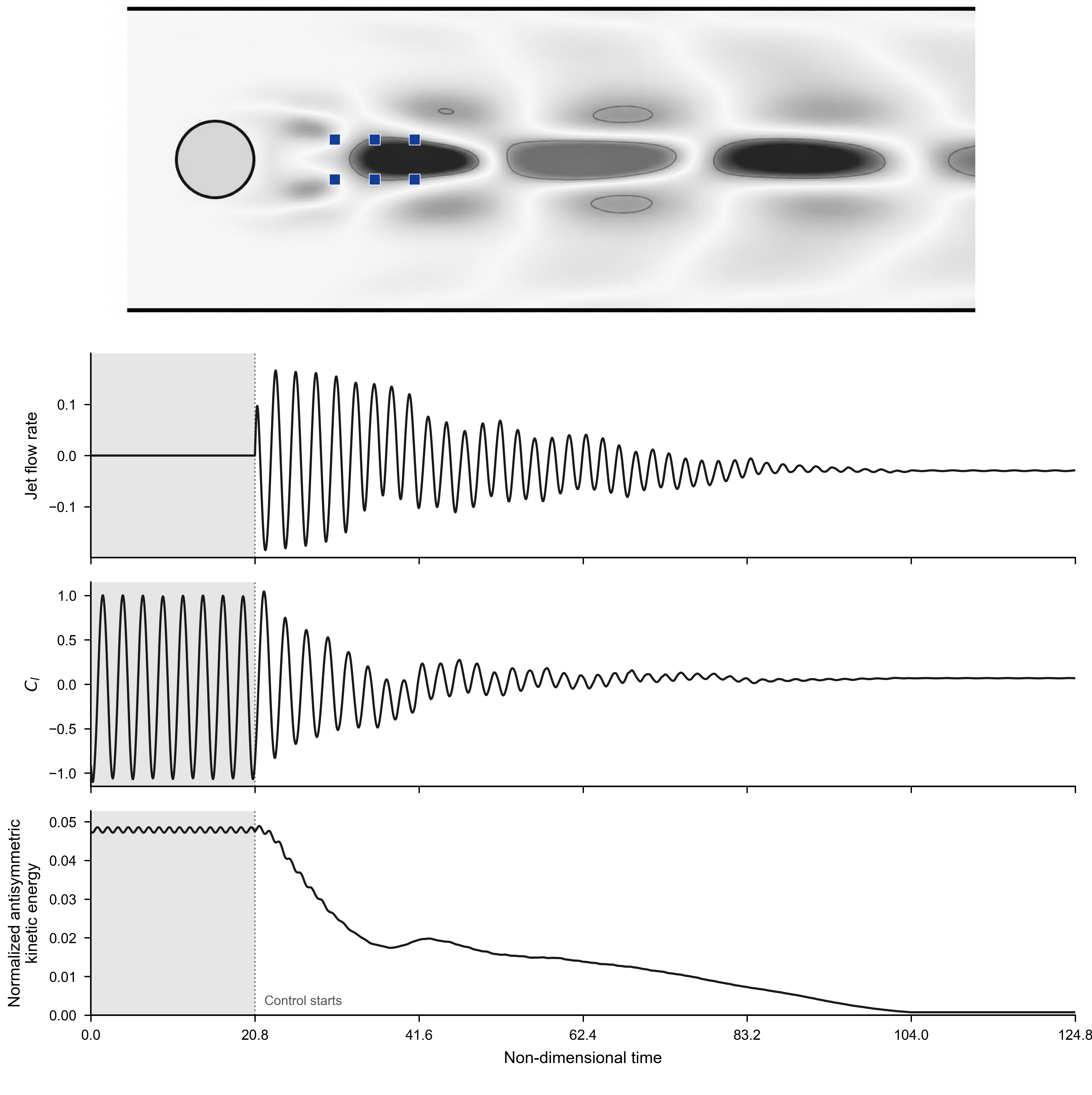}
  \caption{Velocity measurements used in the reward function and the resulting control response.
    Upper schematic: squares mark the six reward-evaluation measurements at
    $(x-x_c)/D=1.5,\,2.0,\,2.5$ and $(y-y_c)/D=\pm0.25$;
    these differ from the policy-observation probes.
    Lower panels show the time
    histories of jet flow rate $q_j$, lift coefficient, and normalized
    antisymmetric kinetic energy $E_A^*$, from top to bottom, over
    $0\leq t\leq124.8$.
    The grey shading denotes the uncontrolled interval, and the vertical
    dotted line marks control activation at $t=20.8$.
    Time $t$ is non-dimensional.}
  \label{fig:probe-loss-control}
\end{fwfigure}

With the velocity-based reward, control begins at $t=20.8$
(figure~\ref{fig:probe-loss-control}).  Large jet flow rates around the
first lift peak and trough give way to progressively weaker actuation as
the lift oscillations decay.  This sequence resembles that obtained with
the force-coefficient reward in \S\ref{sec:complete-and-persistent} and
reported for DRL control of confined wakes by \citet{Li2022Confined}.
Meanwhile, $E_A^*$ falls sharply, rises slightly and then continues to
decline.  Late in the control process, small jet-flow-rate variations
accompany weak lift fluctuations and near-zero $E_A^*$. The
lift-fluctuation suppression reaches approximately $99.8\%$
(table~\ref{tab:design-performance}), comparable to the $99.6\%$ obtained
with the default force-coefficient reward in \S\ref{sec:control-performance}.

With the velocity-based reward, $E_A^*$ takes longer to approach zero
than with the force-coefficient reward (figures~\ref{fig:probe-loss-control}
and~\ref{fig:complete-suppression-energy}). Here, a reward based on local
squared velocity deviations from the prescribed reference values produces
substantial shedding suppression without force coefficients, an SFD base
flow or a DMD growth-rate term.

The squared-deviation loss penalises both mean-flow departure and local
velocity fluctuations. Let $u_i(t)=u(\bfx_i,t)$ and $v_i(t)=v(\bfx_i,t)$,
with $u_i'=u_i-\overline{u_i}$ and $v_i'=v_i-\overline{v_i}$, where an
overbar denotes a temporal average over the same interval. For the fixed
reference values in equation~\eqref{eq:loss-probe},
\begin{equation}\label{eq:probe-loss-mean-variance}
  \overline{\ell_{\mathrm{probe}}}
  =\frac{1}{2N_{\ell}}\sum_{i=1}^{N_{\ell}}
  \left[
    \bigl(\overline{u_i}-u_i^*\bigr)^2
    +\bigl(\overline{v_i}-v_i^*\bigr)^2
    +\overline{u_i'^2}+\overline{v_i'^2}
  \right].
\end{equation}
Thus, the reward contains a direct penalty on local fluctuations without
requiring a separate instability term. The reductions in lift fluctuations
and $E_A^*$ show that optimising this local objective also suppresses
the wake beyond the six reward-evaluation sites.

Using reference values from uncontrolled-flow averages provides a simpler
reward design for differentiable policy optimisation in this wake,
without an SFD base-flow computation or a DMD growth-rate estimate. Alongside
the reward-design findings of \citet{Li2022Confined}, it suggests that
reward requirements may reflect both the flow physics and the policy
optimisation method. We next examine whether the measurements supplied
to the policy can also be simplified.

\subsubsection{Effect of probe configuration}
\label{sec:probe-configuration}

Reducing the number of measurement sites and placing sensors on the body
surface would ease experimental deployment, where access to the wake is
restricted \citep{Wang2024Dynamic}. DRL studies have advanced sparse sensing
through informed probe placement and measurement and action histories
(table~\ref{tab:observation-design-literature}). Here we examine the
observations that support suppression within the present differentiable
optimisation framework. With the default conventional force-coefficient
reward and instantaneous measurements as the sole policy inputs, we assess
how probe number, location and measured quantity affect suppression.

\begin{fwtable}[tbp]
  \centering
  \captionsetup{hypcap=false}
  \caption{Control performance with reduced observations, with
    and without additional observation design. Each row specifies its reference
    configuration. $\Delta\Cd$ denotes drag reduction and $\sigma(\Cl)$ the lift
    standard deviation. Percentages express differences relative to each
    row's reference. Lower $\Delta\Cd$ or higher $\sigma(\Cl)$ indicates poorer
    control. A dash denotes an unreported comparison or design.}
  \label{tab:observation-design-literature}
  {\footnotesize
  \setlength{\tabcolsep}{3pt}
  \renewcommand{\arraystretch}{1.05}
  \begin{tabular}{@{}>{\raggedright\arraybackslash}p{0.23\linewidth}>{\raggedright\arraybackslash}p{0.14\linewidth}>{\raggedright\arraybackslash}p{0.20\linewidth}>{\raggedright\arraybackslash}p{0.18\linewidth}>{\raggedright\arraybackslash}p{\dimexpr0.25\linewidth-24pt\relax}@{}}
    \toprule
    Study and reference & Reduced observations & Without additional design & Additional design & With additional design \\
    \midrule
    \citet{Rabault2019}\newline Circular, $\Rey=100$\newline 151 velocity probes
    & 5 / 11 pressure probes
    & $\Delta\Cd$: $\approx32\%$ / $16\%$ lower
    & \textemdash{} & \textemdash{} \\
    \addlinespace[4pt]
    \citet{Paris2021}\newline Circular, $\Rey=120$\newline 12 wake pressure probes
    & 5 wake pressure probes
    & \textemdash{}
    & Probe selection with S-PPO-CMA
    & $\Delta\Cd$: $\approx2.2\%$ lower \\
    \addlinespace[4pt]
    \citet{Li2022Confined}\newline Circular, $\Rey=150$, $D/H=0.25$\newline 86 velocity probes; near-zero shedding energy
    & 10 velocity probes
    & Residual shedding energy (layouts 2--5)
    & Wavemaker-based placement (layout 1)
    & Near-zero shedding energy for $t>20$ \\
    \addlinespace[4pt]
    \citet{Wang2024Dynamic}\newline Circular, $\Rey=100$\newline 147 pressure probes
    & 4 surface pressure probes
    & $\Delta\Cd$: $\approx20\%$ lower; $\sigma(\Cl)$: $\approx93\%$ higher
    & Dynamic features from pressure and action histories
    & $\Delta\Cd$: gap $\approx0\%$; reduced lift fluctuations \\
    \addlinespace[4pt]
    \citet{Xia2024}\newline Square, $\Rey=100$\newline 64 wake pressure probes
    & 64 rear-surface pressure probes
    & $\Delta\Cd$: $45.1\%$ lower; residual shedding
    & Pressure and action histories
    & $\Delta\Cd$: $4.9\%$ lower; weaker residual shedding \\
    \bottomrule
  \end{tabular}}
\end{fwtable}

The comparisons proceed from 151 velocity probes to one near-wake
velocity probe, then to a surface velocity probe, and finally to pressure at the same surface location (figure~\ref{fig:probe-configuration}).  Each
observation setup uses a separately optimised policy with the default
force-coefficient reward and the common settings in
\S\ref{sec:training}.  Only instantaneous readings are
supplied: two velocity components or one pressure scalar.

The near-wake location follows the structural-sensitivity analysis of
\citet{Giannetti2007}.  The overlap of direct and adjoint modes localises the wavemaker in
two sensitive lobes on either side of the near-wake recirculation region,
where local feedback most strongly changes the unstable-mode eigenvalue.
Using this region to guide probe placement in confined wakes,
\citet{Li2022Confined} obtained near-zero shedding energy with a suitably
distributed ten-probe layout, retaining the suppression achieved with
denser observations; other ten-probe layouts gave appreciably different
results.  Informative placement is thus an established route to sparse
DRL sensing \citep{Paris2021,Li2022Confined}.  The single-probe locations in table~\ref{tab:observation-settings}
allow us to compare near-wake and surface measurements at the same
cross-stream coordinate.

\begin{fwfigure}
  \centering
  \includegraphics[width=\linewidth]{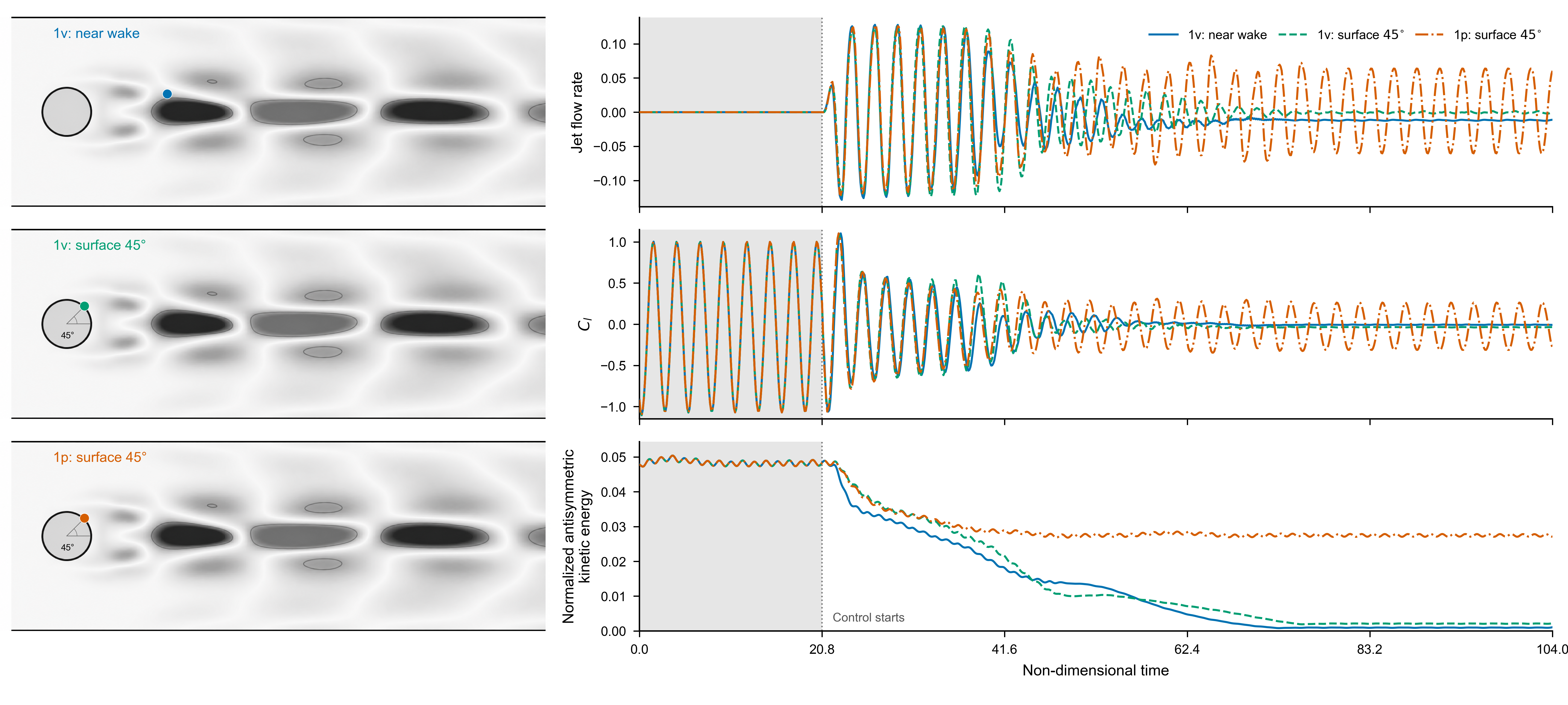}
  \caption{Effect of single-probe observations on control performance.
    Left column, from top to bottom: near-wake velocity probe
    $1.5D$ downstream of the cylinder, surface velocity probe at
    $\theta=45^{\circ}$, and surface pressure probe at the same location.
    Labels 1v and 1p denote one velocity probe (both components) and
    one pressure probe, respectively. All three probes have the same
    $y$ coordinate; the grey wake
    backgrounds are schematic.  Right column, from top to bottom:
    jet flow rate $q_j$, lift coefficient $\Cl$, and normalized antisymmetric
    kinetic energy $E_A^*$, defined in
    equation~\eqref{eq:normalized-antisymmetric-kinetic-energy}, versus
    non-dimensional time $t$.
    Blue solid, green dashed and orange dash-dotted lines denote a
    near-wake velocity probe, a surface velocity probe at
    $\theta=45^{\circ}$ and a surface pressure probe at the same location,
    respectively.  The grey shaded interval precedes control activation;
    the vertical dotted line marks its onset at $t=20.8$.}
  \label{fig:probe-configuration}
\end{fwfigure}

Reducing the 151-probe array to the two instantaneous velocity components
at one near-wake location retains strong suppression
(figure~\ref{fig:probe-configuration}, blue curves).
After activation, the initially large jet oscillations decay, the lift
oscillations become nearly indistinguishable from zero towards the end
of the record, and the normalized antisymmetric kinetic energy approaches
zero. Lift-fluctuation suppression reaches approximately $99.8\%$
(table~\ref{tab:design-performance}).

Moving the velocity probe to $\theta=45^{\circ}$ on the cylinder surface
keeps the measured velocity components $(u,v)$ and the $y$ coordinate fixed.
The surface probe also produces decaying jet oscillations and lift
fluctuations approaching zero (figure~\ref{fig:probe-configuration}, green
curves). Its initial decay of $E_A^*$ is slower than with the near-wake
probe, but continues to a near-zero level by the end of the record.
Lift-fluctuation suppression is again approximately $99.8\%$.
Effective feedback is therefore retained with two instantaneous velocity
components at a single surface location, without direct near-wake sampling
or an observation history.

A single pressure probe on the cylinder surface at $\theta=45^{\circ}$
provides an instantaneous pressure scalar as feedback.  The lift amplitude decreases
markedly relative to the uncontrolled state, and $E_A^*$ falls to a lower
plateau (figure~\ref{fig:probe-configuration}, orange curves).
The lift-fluctuation suppression rate is $72.3\%$, compared with
approximately $99.8\%$ for either velocity observation setting.
The persistent lift oscillations and non-zero antisymmetric energy show that
pressure feedback attenuates the wake fluctuations without fully
suppressing them.

\begin{fwtable}[tbp]
  \centering
  \caption{Wake suppression with simplified rewards or observations.
    Each setting uses a separately optimised policy. Statistics cover the
    final five uncontrolled shedding periods of each evaluation record.
    Lift-fluctuation suppression is relative to the uncontrolled
    $\sigma(\Cl)=0.7390$; $\overline{E_A^*}$ is the mean normalised
    antisymmetric kinetic energy.}
  \label{tab:design-performance}
  \small
  \begin{tabular}{@{}p{0.29\linewidth}p{0.23\linewidth}cc@{}}
    \toprule
    Policy observations & Reward & $\Delta\sigma(\Cl)\,(\%)$ & $\overline{E_A^*}$ \\
    \midrule
    151 velocity probes & Six-point velocity & 99.8 & $7.38\times10^{-4}$ \\
    One near-wake velocity probe & Force coefficients & 99.8 & $9.60\times10^{-4}$ \\
    One surface velocity probe & Force coefficients & 99.8 & $2.12\times10^{-3}$ \\
    One surface pressure probe & Force coefficients & 72.3 & $2.75\times10^{-2}$ \\
    \bottomrule
  \end{tabular}
\end{fwtable}

The comparisons separate two routes to simpler sensing: reducing the
number of velocity probes retains strong suppression, and moving the
remaining probe to the surface retains effective feedback without direct
wake measurements. At the same surface location, two instantaneous
velocity components support nearly complete suppression, whereas one
instantaneous pressure scalar leaves residual oscillations. Related DRL
studies show how temporal information can improve control: \citet{Wang2024Dynamic} constructed dynamic features from
pressure and action histories for circular-cylinder control with one
surface pressure sensor.  In square-cylinder control, \citet{Xia2024}
obtained partial shedding suppression using current surface pressures and
the latest action.  Longer measurement and action histories improved drag
reduction, but performance still deteriorated when only one surface
pressure sensor was retained.

\citet{Li2022Confined} used structural sensitivity to explain why probes
covering the wavemaker provide effective observations for DRL control.
\citet{Wang2024Dynamic} linked difficulties in policy learning from
surface pressure to insufficient flow-state information and used dynamic
features to improve the observations. \citet{Xia2024} related the
limitations of surface measurements to missing streamwise information
and convective delays, and incorporated measurement and action histories
to address partial observability.
These explanations relate the flow information available to the policy
to the difficulty of learning effective control. In the present study,
nearly complete suppression is retained using local instantaneous velocity
measurements at the tested near-wake or surface location, without temporal
histories or a distributed wake-probe array. During optimisation, gradients
through the simulated flow evolution convey how actuation affects the
control loss (\S\ref{sec:gradients}), although the policy input remains local.
This distinction between information used to optimise the policy and
measurements supplied to it suggests that observation requirements may
depend partly on the policy optimisation method, as well as on the wake
dynamics.

These separate comparisons show that the reward and policy observations
can each be simplified to some extent while retaining strong shedding
suppression in the tested wake. The contrast between velocity and pressure
observations at the same surface location also shows that the effectiveness
of sparse sensing depends on the measured quantity.

\section{Conclusions}\label{sec:conclusions}

We have applied differentiable simulation to closed-loop policy
optimisation for synthetic-jet control of a two-dimensional confined
circular-cylinder wake at $\Rey=100$. Differentiable jet velocity boundary
conditions connect the neural-network policy to the flow simulation,
allowing control-loss gradients to propagate to the policy parameters.
The study evaluates training interactions and control performance, examines how
the policy sustains suppression, and separately assesses simpler reward
and observation designs.

With the default 151-probe velocity observations
and a conventional force-coefficient reward, the policy achieves
approximately $8\%$ drag reduction and $99.6\%$ lift-fluctuation suppression.
Training loss decreases towards a stable value over eight policy updates,
with a cumulative interaction budget of about $78$ uncontrolled shedding
periods. These control improvements are obtained with fewer training
interactions than reported in the two DRL studies in
table~\ref{tab:drl-comparison}.

The initially large jet adjustments weaken as wake oscillations decay,
and small, time-varying feedback sustains the nearly symmetric,
low-fluctuation wake. Constant actuation at the same mean jet flow rate
does not maintain suppression. At three different starting phases,
the same policy produces different initial outputs, followed by similar
jet and lift responses with decaying oscillations. Wake asymmetry follows
a decreasing trend consistent with the baseline control response at all
three phases. The policy also restores suppression after either of the
two upstream disturbances is removed, without retraining. These control
responses are learned with the conventional force-coefficient reward based
on drag and lift, without additional reward terms based on flow stability.

Separate comparisons show that reward and observation designs can each
be simplified while retaining strong suppression. With the default
151-probe observations, a simple velocity-based reward uses only squared
velocity deviations at six downstream locations, referenced to the
symmetrised mean uncontrolled wake. On time averaging, these squared
deviations penalise both mean-flow departures and local velocity
fluctuations. Although the transient towards a symmetric wake is longer
than with the force-coefficient reward, strong suppression is achieved
without force coefficients, base-flow calculations or instability
penalties. In the observation comparison, the conventional
force-coefficient reward is retained. Policies optimised separately with
a single velocity probe in the near wake or on the cylinder surface both
achieve nearly complete suppression. Each policy receives only the two
instantaneous velocity components, without measurement or action
histories. A single pressure value at the same surface location attenuates
the oscillations but does not eliminate them. Local instantaneous
velocity observations therefore suffice at both tested locations, while
the measured quantity affects the suppression achieved.

These findings establish differentiable simulation as an effective route
to wake-control policy optimisation, with reward and sensing
simplification demonstrated separately in the present configuration.
Together with related DRL findings, they also suggest that reward and
observation requirements may depend partly on the policy optimisation
method. Further studies of other simple reward functions and single
probes at other cylinder surface locations, together with extensions to
higher Reynolds numbers and three-dimensional wakes, could advance
efficient closed-loop control strategies for more complex flows.

\section*{Acknowledgements}
This work was supported by the National Key R\&D Program of China
(Grant No. 2024YFB4207000) and the National Natural Science Foundation of
China (Grant No. U25A20372).

\section*{Declaration of interests}
The authors report no conflict of interest.

\section*{Data and code availability}
The repository for the policy optimisation framework is \url{https://github.com/alanZee/D-AFC}.

\section*{CRediT authorship contribution statement}
\textbf{Siwen Zhang:} Conceptualization, Methodology, Software, Validation,
Data analysis, Data curation, Visualization, Writing -- original draft \& editing.\par
\textbf{Xizeng Zhao:} Supervision, Resources, Project
administration, Funding acquisition, Writing -- review \& editing.\par
\textbf{Zhouteng Ye:} Conceptualization, Methodology, Validation,
Writing -- review \& editing.\par
\textbf{Yulin Xie:} Conceptualization, Writing -- review \& editing.\par
\textbf{Zesheng Yao:} Conceptualization, Data analysis.\par
All authors discussed the results and commented on the manuscript.


\appendix
\renewcommand*{\theHsection}{appendix.\Alph{section}}
\section{Dynamic mode decomposition}
\label{app:dmd}

DMD approximates the evolution between consecutive flow snapshots by a
linear mapping. We estimate its spectrum
using a projected matrix constructed from the singular value decomposition
of the snapshot sequence.

Each DMD estimate uses a window of approximately $1.5$ uncontrolled shedding
periods. The window contains 20 flow-field snapshots separated by $\Delta t_{\mathrm{snap}}=0.2$ (40 solver steps). Its first-to-last snapshot span is therefore
$19\Delta t_{\mathrm{snap}}=3.8$ in the solver's non-dimensional time.

For a discrete DMD eigenvalue $\lambda_i$, the
growth rate is $\operatorname{Re}(\log\lambda_i)/\Delta t_{\mathrm{snap}}$ and its frequency
is $\operatorname{Im}(\log\lambda_i)/(2\pi\Delta t_{\mathrm{snap}})$.  Both use the solver's
non-dimensional time.  The leading mode is selected by the largest
estimated growth rate; its vorticity distribution is used to compare
spatial structure in figure~\ref{fig:leading-dmd-modes}.

\section{Upstream disturbance}
\label{app:upstream-disturbance}

The upstream disturbance is an external force
$\mathbf{F}(x,y)\xi(t)$ with a Gaussian spatial profile
\citep{Li2022Confined}:
\begin{equation}\label{eq:upstream-forcing}
  \mathbf{F}(x,y)=A
  \exp\!\left[-\frac{(x-x_0)^2}{2\sigma_F^2}\right]
  \exp\!\left[-\frac{(y-y_0)^2}{2\sigma_F^2}\right]
  (1,1)^{\mathsf{T}},
\end{equation}
where $\sigma_F=0.1$ and $(x_0,y_0)=(-1.5,0)$ in the
cylinder-centred coordinates.  The random scaling factor $\xi(t)$
has standard deviation $\sigma_\xi=0.1$.

\bibliographystyle{elsarticle-harv}
\bibliography{refs}

\end{document}